%
\let\includefigures=\iftrue
%
%
%
%
%
\input harvmac
\input rotate
\input epsf
\input xyv2
\noblackbox
\includefigures
\message{If you do not have epsf.tex (to include figures),}
\message{change the option at the top of the tex file.}
\def\figin{\epsfcheck\figin}\def\figins{\epsfcheck\figins}
\def\epsfcheck{\ifx\epsfbox\UnDeFiNeD
\message{(NO epsf.tex, FIGURES WILL BE IGNORED)}
\gdef\figin##1{\vskip2in}\gdef\figins##1{\hskip.5in}
\else\message{(FIGURES WILL BE INCLUDED)}%
\gdef\figin##1{##1}\gdef\figins##1{##1}\fi}
\def\DefWarn#1{}

\def\figinsert{\goodbreak\midinsert}
\def\ifig#1#2#3{\DefWarn#1\xdef#1{fig.~\the\figno}
\writedef{#1\leftbracket fig.\noexpand~\the\figno}%
\figinsert\figin{\centerline{#3}}\medskip\centerline{\vbox{\baselineskip12pt
\advance\hsize by -1truein\noindent\footnotefont{\bf
Fig.~\the\figno:} #2}}
\bigskip\endinsert\global\advance\figno by1}
\else
\def\ifig#1#2#3{\xdef#1{fig.~\the\figno}
\writedef{#1\leftbracket fig.\noexpand~\the\figno}%
\global\advance\figno by1} \fi
\def\Title#1#2{\rightline{#1}\ifx\answ\bigans\nopagenumbers\pageno0
\else\pageno1\vskip.5in\fi \centerline{\titlefont #2}\vskip .3in}
\font\caps=cmcsc10

\def\yboxit#1#2{\vbox{\hrule height #1 \hbox{\vrule width #1
\vbox{#2}\vrule width #1 }\hrule height #1 }}
\def\fillbox#1{\hbox to #1{\vbox to #1{\vfil}\hfil}}
\def\ybox{{\lower 1.3pt \yboxit{0.4pt}{\fillbox{8pt}}\hskip-0.2pt}}

\def\rightarrowbox#1#2{
  \setbox1=\hbox{\kern#1{${ #2}$}\kern#1}
  \,\vbox{\offinterlineskip\hbox to\wd1{\hfil\copy1\hfil}
    \kern 3pt\hbox to\wd1{\rightarrowfill}}}

\def\half{{1\over 2}}

\def\bra#1{{\langle}#1|}
\def\ket#1{|#1\rangle}

\def\CL{{\cal L}}

\def\CN{{\cal N}}
\def\CO{{\cal O}}

\def\tilde{\widetilde}

           \def\CO{{\cal O}} 
   
\def\CL{{\cal L}}   
 \def\CR{{\cal R}}  
   
\def\CN{{\cal N}}


\def\dj{\hbox{d\kern-0.347em \vrule width 0.3em height 1.252ex depth
-1.21ex \kern 0.051em}}

\def\half{{1\over 2}\,}

\def\ket{\rangle}
\def\bra{\langle}

\def\pt{\partial}

\def\Dirac{\,\raise.15ex\hbox{/}\mkern-13.5mu D}
\def\dirac{\,\raise.15ex\hbox{/}\kern-.57em \partial}
\def\aslash{\,\raise.15ex\hbox{/}\mkern-13.5mu A}

\def\shalf{{\ifinner {\textstyle {1 \over 2}}\else {1 \over 2} \fi}}
\def\sshalf{{\ifinner {\scriptstyle {1 \over 2}}\else {1 \over 2} \fi}}
\def\sfourth{{\ifinner {\textstyle {1 \over 4}}\else {1 \over 4} \fi}}
\def\sthreehalfs{{\ifinner {\textstyle {3 \over 2}}\else {3 \over 2} \fi}}
\def\sdhalfs{{\ifinner {\textstyle {d \over 2}}\else {d \over 2} \fi}}
\def\sdmtwohalfs{{\ifinner {\textstyle {d-2 \over 2}}\else {d-2 \over 2} \fi}}
\def\sdmasonehalfs{{\ifinner {\textstyle {d+1 \over 2}}\else {d+1 \over 2} \fi}}
\def\sdmasthreehalfs{{\ifinner {\textstyle {d+3 \over 2}}\else {d+3 \over 2} \fi}}
\def\sdmastwohalfs{{\ifinner {\textstyle {d+2 \over 2}}\else {d+2 \over 2} \fi}}
\def\te{{\rm t}}


\lref\cdl{
 S.~R.~Coleman and F.~De Luccia,
  ``Gravitational Effects On And Of Vacuum Decay,''
  Phys.\ Rev.\  D {\bf 21}, 3305 (1980).
  
   L.~F.~Abbott and S.~R.~Coleman,
  ``The Collapse Of An Anti-De Sitter Bubble,''
  Nucl.\ Phys.\  B {\bf 259}, 170 (1985).
}

\lref\us{
  J.~L.~F.~Barb\'on and E.~Rabinovici,
  ``Holography of AdS vacuum bubbles,''
  JHEP {\bf 1004}, 123 (2010)
  [arXiv:1003.4966 [hep-th]].}

\lref\MaldacenaUN{
  J.~Maldacena,
  ``Vacuum decay into Anti de Sitter space,''
  arXiv:1012.0274 [hep-th].
}

 \lref\adscft{
  J.~M.~Maldacena,
  ``The large N limit of superconformal field theories and supergravity,''
  Adv.\ Theor.\ Math.\ Phys.\  {\bf 2}, 231 (1998)
  [Int.\ J.\ Theor.\ Phys.\  {\bf 38}, 1113 (1999)]
  [arXiv:hep-th/9711200].
 S.~S.~Gubser, I.~R.~Klebanov and A.~M.~Polyakov,
  ``Gauge theory correlators from non-critical string theory,''
  Phys.\ Lett.\  B {\bf 428}, 105 (1998)
  [arXiv:hep-th/9802109].
 E.~Witten,
  ``Anti-de Sitter space and holography,''
  Adv.\ Theor.\ Math.\ Phys.\  {\bf 2}, 253 (1998)
  [arXiv:hep-th/9802150].
  }

\lref\SusskindIF{
  L.~Susskind, L.~Thorlacius and J.~Uglum,
  ``The Stretched Horizon And Black Hole Complementarity,''
  Phys.\ Rev.\  D {\bf 48}, 3743 (1993)
  [arXiv:hep-th/9306069].
}

\lref\mogollon{
S.~Hawking, J.~M.~Maldacena and A.~Strominger,
  ``DeSitter entropy, quantum entanglement and AdS/CFT,''
  JHEP {\bf 0105}, 001 (2001)
  [arXiv:hep-th/0002145].

A.~Buchel and A.~A.~Tseytlin,
   ``Curved space resolution of singularity of fractional D3-branes on
  conifold,''
  Phys.\ Rev.\  D {\bf 65}, 085019 (2002)
  [arXiv:hep-th/0111017].

  A.~Buchel, P.~Langfelder and J.~Walcher,
  ``Does the tachyon matter?,''
  Annals Phys.\  {\bf 302}, 78 (2002)
  [arXiv:hep-th/0207235].

  A.~Buchel,
  ``Gauge / gravity correspondence in accelerating universe,''
  Phys.\ Rev.\  D {\bf 65}, 125015 (2002)
  [arXiv:hep-th/0203041].

A.~Buchel, P.~Langfelder and J.~Walcher,
  ``On time-dependent backgrounds in supergravity and string theory,''
  Phys.\ Rev.\  D {\bf 67}, 024011 (2003)
  [arXiv:hep-th/0207214].
  T.~Hirayama,
  ``A holographic dual of CFT with flavor on de Sitter space,''
  JHEP {\bf 0606}, 013 (2006)
  [arXiv:hep-th/0602258].
M.~Alishahiha, A.~Karch, E.~Silverstein and D.~Tong,
  ``The dS/dS correspondence,''
  AIP Conf.\ Proc.\  {\bf 743}, 393 (2005)
  [arXiv:hep-th/0407125].

  A.~Buchel,
  ``Inflation on the resolved warped deformed conifold,''
  Phys.\ Rev.\  D {\bf 74}, 046009 (2006)
  [arXiv:hep-th/0601013].

  O.~Aharony, M.~Fabinger, G.~T.~Horowitz and E.~Silverstein,
  ``Clean time-dependent string backgrounds from bubble baths,''
  JHEP {\bf 0207}, 007 (2002)
  [arXiv:hep-th/0204158].

  V.~Balasubramanian and S.~F.~Ross,
  ``The dual of nothing,''
  Phys.\ Rev.\  D {\bf 66}, 086002 (2002)
  [arXiv:hep-th/0205290].

  S.~F.~Ross and G.~Titchener,
  ``Time-dependent spacetimes in AdS/CFT: Bubble and black hole,''
  JHEP {\bf 0502}, 021 (2005)
  [arXiv:hep-th/0411128].

  R.~G.~Cai,
  ``Constant curvature black hole and dual field theory,''
  Phys.\ Lett.\  B {\bf 544}, 176 (2002)
  [arXiv:hep-th/0206223].

  V.~Balasubramanian, K.~Larjo and J.~Simon,
  ``Much ado about nothing,''
  Class.\ Quant.\ Grav.\  {\bf 22}, 4149 (2005)
  [arXiv:hep-th/0502111].

  J.~He and M.~Rozali,
  ``On Bubbles of Nothing in AdS/CFT,''
  JHEP {\bf 0709}, 089 (2007)
  [arXiv:hep-th/0703220].

 J.~A.~Hutasoit, S.~P.~Kumar and J.~Rafferty,
   ``Real time response on dS$_3$: the Topological AdS Black Hole and the
  Bubble,''
  JHEP {\bf 0904}, 063 (2009)
  [arXiv:0902.1658 [hep-th]].

}
\lref\Marolf{
  D.~Marolf, M.~Rangamani and M.~Van Raamsdonk,
  ``Holographic models of de Sitter QFTs,''
  arXiv:1007.3996 [hep-th].}

\lref\HertogH{
  T.~Hertog and G.~T.~Horowitz,
  ``Holographic description of AdS cosmologies,''
  JHEP {\bf 0504}, 005 (2005)
  [arXiv:hep-th/0503071].
   T.~Hertog and G.~T.~Horowitz,
  ``Towards a big crunch dual,''
  JHEP {\bf 0407}, 073 (2004)
  [arXiv:hep-th/0406134].
}
\lref\insightfull{
  G.~Horowitz, A.~Lawrence and E.~Silverstein,
  ``Insightful D-branes,''
  JHEP {\bf 0907}, 057 (2009)
  [arXiv:0904.3922 [hep-th]].}

\lref\hartleh{
  J.~B.~Hartle and S.~W.~Hawking,
  ``Wave Function Of The Universe,''
  Phys.\ Rev.\  D {\bf 28}, 2960 (1983).
}
\lref\Polyakov{
  A.~M.~Polyakov,
  ``Decay of Vacuum Energy,''
  Nucl.\ Phys.\  B {\bf 834}, 316 (2010)
  [arXiv:0912.5503 [hep-th]].}
\lref\nothing{
  E.~Witten,
  ``Instability Of The Kaluza-Klein Vacuum,''
  Nucl.\ Phys.\  B {\bf 195}, 481 (1982).}

\lref\hsus{
  D.~Harlow and L.~Susskind,
  ``Crunches, Hats, and a Conjecture,''
  arXiv:1012.5302 [hep-th].}
\lref\harlow{
  D.~Harlow,
  ``Metastability in Anti de Sitter Space,''
  arXiv:1003.5909 [hep-th].}

\lref\banks{
T.~Banks,
  ``Heretics of the false vacuum: Gravitational effects on and of vacuum decay.
  II,''
  arXiv:hep-th/0211160.
T.~Banks,
  ``Landskepticism or why effective potentials don't count string models,''
  arXiv:hep-th/0412129.

  T.~Banks,
  ``TASI Lectures on Holographic Space-Time, SUSY and Gravitational Effective
  Field Theory,''
  arXiv:1007.4001 [hep-th].
}

\lref\banksc{
  T.~Banks and W.~Fischler,
  ``Space-like singularities and thermalization,''
  arXiv:hep-th/0606260.

  T.~Banks,
  ``Pedagogical notes on black holes, de Sitter space, and bifurcated
  horizons,''
  arXiv:1007.4003 [hep-th].
}

\lref\bt{
  P.~Breitenlohner and D.~Z.~Freedman,
  ``Stability In Gauged Extended Supergravity,''
  Annals Phys.\  {\bf 144}, 249 (1982).

  P.~Breitenlohner and D.~Z.~Freedman,
  ``Positive Energy In Anti-De Sitter Backgrounds And Gauged Extended
  Supergravity,''
  Phys.\ Lett.\  B {\bf 115}, 197 (1982).
}
  
\lref\magan{
  J.~L.~F.~Barb\'on and J.~Mart\'{\i}nez-Mag\'an,
  ``Spontaneous fragmentation of topological black holes,''
  JHEP {\bf 1008}, 031 (2010)
  [arXiv:1005.4439 [hep-th]].
}

\lref\guthblau{
S.~K.~Blau, E.~I.~Guendelman and A.~H.~Guth,
  ``The Dynamics of False Vacuum Bubbles,''
  Phys.\ Rev.\  D {\bf 35}, 1747 (1987).
  G.~L.~Alberghi, D.~A.~Lowe and M.~Trodden,
  ``Charged false vacuum bubbles and the AdS/CFT correspondence,''
  JHEP {\bf 9907}, 020 (1999)
  [arXiv:hep-th/9906047].
  B.~Freivogel, V.~E.~Hubeny, A.~Maloney, R.~C.~Myers, M.~Rangamani and S.~Shenker,
  ``Inflation in AdS/CFT,''
  JHEP {\bf 0603}, 007 (2006)
  [arXiv:hep-th/0510046].
}

\lref\israel{
W.~Israel,
  ``Singular hypersurfaces and thin shells in general relativity,''
  Nuovo Cim.\  B {\bf 44S10}, 1 (1966)
  [Erratum-ibid.\  B {\bf 48}, 463 (1967\ NUCIA,B44,1.1966)].
}

\lref\seibwit{
N.~Seiberg and E.~Witten,
  ``The D1/D5 system and singular CFT,''
  JHEP {\bf 9904}, 017 (1999)
  [arXiv:hep-th/9903224].
}

  \lref\alberghi{
  G.~L.~Alberghi, D.~A.~Lowe and M.~Trodden,
  ``Charged false vacuum bubbles and the AdS/CFT correspondence,''
  JHEP {\bf 9907}, 020 (1999)
  [arXiv:hep-th/9906047].
  }
  
\lref\dtrace{
E.~Witten,
  ``Multi-trace operators, boundary conditions, and AdS/CFT correspondence,''
  arXiv:hep-th/0112258.
  M.~Berkooz, A.~Sever and A.~Shomer,
  ``Double-trace deformations, boundary conditions and spacetime
  singularities,''
  JHEP {\bf 0205}, 034 (2002)
  [arXiv:hep-th/0112264].
}

\lref\fubini{
S.~Fubini, ``A New Approach To Conformal Invariant Field Theories,''
Nuovo Cim.\ A {\bf 34}, 521 (1976).}

\lref\rabroll{
V.~Asnin, E.~Rabinovici and M.~Smolkin,
  ``On rolling, tunneling and decaying in some large N vector models,''
  JHEP {\bf 0908}, 001 (2009)
  [arXiv:0905.3526 [hep-th]].
  }
  
  \lref\ofer{
  O.~Aharony, B.~Kol and S.~Yankielowicz,
  ``On exactly marginal deformations of N = 4 SYM and type IIB  supergravity on
  ${\rm AdS}_5 \times S^5$,''
  JHEP {\bf 0206}, 039 (2002)
  [arXiv:hep-th/0205090].
}

\lref\craps{
A.~Bernamonti and B.~Craps,
  ``D-Brane Potentials from Multi-Trace Deformations in AdS/CFT,''
  JHEP {\bf 0908}, 112 (2009)
  [arXiv:0907.0889 [hep-th]].
}

\lref\eli{
S.~Elitzur, A.~Giveon, M.~Porrati and E.~Rabinovici,
  ``Multitrace deformations of vector and adjoint theories and their
  holographic duals,''
  JHEP {\bf 0602}, 006 (2006)
  [arXiv:hep-th/0511061].  {\it ibid} Nucl.\ Phys.\ Proc.\ Suppl.\  {\bf 171}, 231 (2007).
  
}

 \lref\turok{
 B.~Craps, T.~Hertog and N.~Turok,
  ``Quantum Resolution of Cosmological Singularities using AdS/CFT,''
  arXiv:0712.4180 [hep-th].
 
  N.~Turok, B.~Craps and T.~Hertog,
  ``From Big Crunch to Big Bang with AdS/CFT,''
  arXiv:0711.1824 [hep-th].
}

\lref\tassos{
S.~de Haro and A.~C.~Petkou,
  ``Instantons and conformal holography,''
  JHEP {\bf 0612}, 076 (2006)
  [arXiv:hep-th/0606276].

 S.~de Haro, I.~Papadimitriou and A.~C.~Petkou,
  ``Conformally coupled scalars, instantons and vacuum instability in
  AdS(4),''
  Phys.\ Rev.\ Lett.\  {\bf 98}, 231601 (2007)
  [arXiv:hep-th/0611315].

 I.~Papadimitriou,
  ``Multi-Trace Deformations in AdS/CFT: Exploring the Vacuum Structure of
  the Deformed CFT,''
  JHEP {\bf 0705}, 075 (2007)
  [arXiv:hep-th/0703152].
  }

\lref\hertogu{
L.~Battarra and T.~Hertog,
  ``Particle Production near an AdS Crunch,''
  JHEP {\bf 1012}, 017 (2010)
  [arXiv:1009.0992 [hep-th]].
  }

\lref\kofman{
  L.~Kofman,
  ``Tachyonic preheating,''
  arXiv:hep-ph/0107280.
  }


\line{\hfill IFT-UAM/CSIC-11-06}

\vskip 0.7cm

\Title{\vbox{\baselineskip 12pt\hbox{}
 }}
{\vbox {\centerline{AdS Crunches, CFT Falls }
\vskip10pt
\centerline{ And}
\vskip10pt
\centerline{Cosmological Complementarity}
}}

\vskip 0.5cm

\centerline{$\quad$ {
{\caps Jos\'e L.F. Barb\'on}$^\dagger$
 {\caps and}
{\caps Eliezer Rabinovici}$^\star$
}}
\vskip1.0cm

\centerline{{\sl  $^\dagger$ Instituto de F\'{\i}sica Te\'orica IFT UAM/CSIC }}
\centerline{{\sl  C-XVI,
 UAM, Cantoblanco 28049. Madrid, Spain }}
\centerline{{\tt jose.barbon@uam.es}}


\centerline{{\sl $^\star$
Racah Institute of Physics, The Hebrew University }}
\centerline{{\sl Jerusalem 91904, Israel}}
\centerline{{\tt eliezer@vms.huji.ac.il}}

\vskip0.2cm

\centerline{\bf ABSTRACT}

 \vskip 0.3cm

 \noindent

We discuss aspects of the holographic description of  crunching AdS cosmologies. We argue that crunching FRW models with hyperbolic spatial sections  are dual to  semiclassical condensates in deformed de Sitter CFTs. De Sitter-invariant  condensates with a sharply defined energy scale are induced by effective negative-definite relevant or marginal operators, which may or may not destabilize the CFT. We find this result by explicitly constructing a `complementarity map' for this model, given by a conformal transformation of the de Sitter CFT into a static time-frame, which reveals the crunch as  an infinite potential-energy fall in finite time.  We show that, quite generically, the crunch is associated to a finite mass black hole if the  de Sitter $O(d,1)$-invariance is  an accidental IR symmetry, broken down  to $U(1)\times O(d)$ in the UV. Any such regularization cuts off the eternity of de Sitter space-time.  Equivalently, the dimension of the Hilbert space propagating into the crunch is finite only when de Sitter is  not  eternal.

\vskip 0.2cm

\Date{February  2011}

\vfill





\baselineskip=15pt

\newsec{Introduction}

\noindent

The resolution of cosmological singularities has been a permanent fixture in the `to do' list of string theory since its quantum-gravitational interpretation was launched. Successful singularity resolutions by various instances of `stringy geometry'   always apply to time-like singularities that may be regarded as `impurities' in space . These singularities are resolved by a  refinement in the quantum description of the impurity, often by identifying additional light degrees of freedom supported at the singularity locus. 

Space-like singularities looking like {\it bangs} and {\it crunches} in General Relativity have so far resisted close scrutiny. At the most naive level,  they represent a challenge to the very notion of Hamiltonian time evolution. In the case of space-like singularities censored by black hole horizons, they  posed a famous historic challenge  to S-matrix unitarity. The advent of notions such as black-hole complementarity,  holography and the AdS/CFT correspondence  have resulted in a conceptual framework in which black hole singularities should be resolved by a refinement of the hole's quantum description, albeit non-perturbative  and highly non-local in this case. 

On general grounds, any problem that can be successfully embedded  into an AdS/CFT environment should admit a honest Hamiltonian verdict. In this sense, the occurrence of crunches in the interior of AdS black holes  poses no serious threat to the CFT Hamiltonian description. The black hole and everything inside is described by a finite-dimensional subspace of the full Hilbert space, and thus the crunch must be `deconstructed' within this class of states, even if the details of such a deconstruction remain largely unknown. 

Potentially more serious is the situation where a crunch engulfs a globally defined, asymptotically AdS space-time, for then  the whole CFT Hilbert space seems destined to the crunch. Such a radical situation occurs in the future development of Coleman--de Luccia  (CdL) bubbles nucleated inside a false AdS vacuum \refs\cdl. The interior of the bubble contains a crunching
cosmology which eventually engulfs the whole AdS space-time, right up to the boundary, not to mention the complicated pattern of multi-bubble nucleation and collision. This poses a `clear and present danger' for the CFT Hamiltonian picture.  

In this note we discuss various aspects of this conundrum. We first review the construction of  highly symmetric  bubble-like backgrounds with a crunchy destiny and AdS asymptotic behavior. These backgrounds include the CdL single-bubble configurations but can be much more general, and are characterized by an exact $O(d,1)$ symmetry. On the dual holographic side this symmetry can be realized by specifying the CFT states as living on an eternal $d$-dimensional de Sitter space-time. 

A central observation in this paper is the identification of a simple `complementarity map' which relates the  formulation of the CFT in de Sitter space-time to another description appropriate for an `observer' who falls into the crunch in finite time.  The large amount of symmetry determines this map to be a  conformal transformation to the same CFT defined on a static Einstein space-time.  Using results from \refs\us\ we argue that the  crunching states are seen in this frame as   infinite negative-energy falls, their precise nature depending on whether the condensate was stable or unstable in the  de Sitter-frame description. We discuss the implications of this statement for the interpretation in terms of a `cosmological complementarity', using in particular a regularization which strictly reverts the model into a standard case of black hole complementarity \refs\SusskindIF.

\newsec{AdS crunches and their dS duals}

\noindent

In this section we  review the basic issues arising  in the construction of  CFT duals of AdS crunching cosmologies. We begin
with a description of the relevant geometries and we subsequently introduce  the corresponding dual CFT structures. Many details of relevance to these constructions can be found in \refs{\HertogH,\eli,\turok, \mogollon,\Marolf,\MaldacenaUN}. 

\subsec{The crunches}

\noindent 

We shall refer to  `AdS-crunches' as a particular class  of 
 FRW cosmologies with $O(d,1)$-invariant spatial sections (i.e. $d$-hyperboloids ${\bf H}^d$), 
\eqn\crunchin{
ds^2_{\rm FRW} = -d{\rm t}^2 + G({\rm t}) \,ds^2_{{\bf H}^d}\;.
}
The profile function $G(\te)$  solves Einstein's equations with negative cosmological constant and a generic $O(d,1)$-invariant matter distribution modeled by a set of fields $\varphi({\rm t})$ depending only on the FRW time coordinate $\te$.  In general, for smooth initial conditions at some fixed time $\te=\te_0$, this    space-time has  curvature singularities both in the future (crunch) and the past (bang). If the matter contribution is small, the metric is close to an exact AdS$_{d+1}$ in FRW parametrization, corresponding to $G(\te)=\sin^2 (\te)$,   
at least for a long time.  In the pure AdS case, the points ${\rm t}=0, \pi$ are only coordinate singularities signaling the  Killing horizons associated to the hyperbolic sections becoming null at this locus.  

The behavior of the FRW patch in pure AdS suggests that a crunching cosmology of type \crunchin\ could be given initial smooth data by matching across a zero of the function $G(\te)$, tuned with a locally static matter distribution. This is precisely the case if the FRW cosmology is regarded as the interior of an expanding bubble, in a generalization of the classic work by Coleman and de Luccia \refs\cdl.  It is convenient to parametrize such backgrounds in terms of the Euclidean versions with $O(d+1)$ isometries.  Let us consider the metric
\eqn\boladef{
ds^2_{\rm Ball} = d\rho^2 +  F(\rho)\,d\Omega_d^2\;,
}
satisfying  the field Equations with an $O(d+1)$-symmetric matter distribution $\varphi(\rho)$. We term it `the ball' on account of its $O(d+1)$ symmetry, even if it may be non-compact in general.        Smoothness at the center of the ball  requires 
  $F(\rho) \approx \rho^2$ and $\varphi(\rho) \approx \varphi_0 + \shalf \varphi''_0 \rho^2$ as $\rho\rightarrow 0$.

Writing  
$
d\Omega_d^2 = d\theta^2 + \cos^2(\theta)\,d\Omega_{d-1}^2
$, we generate a Lorentz-signature metric with $O(d,1)$ symmetry by the analytic continuation  $\theta = i\tau $. We call this metric `the bubble': 
\eqn\adsds{
ds^2_{\rm Bubble} = d\rho^2 + F(\rho) \,\left(-d\tau^2 +  \cosh^2 (\tau) \,d\Omega_{d-1}^2 \right) = d\rho^2 + F(\rho)\,ds^2_{{\rm dS}_d}\;, 
}
where the group $O(d,1)$ acts on global de Sitter sections dS$_d$. By construction, the matter fields are de Sitter-invariant functions $\varphi(\rho)$, so all features of the metric and matter fields expand like a de Sitter space-time, i.e. we have a generalized notion of an `expanding bubble'. This bubble background is time-symmetric around $\tau=0$, where it can be formally matched to the Euclidean $O(d+1)$-invariant `ball'. Therefore, we may interpret this construction as a time-symmetric cosmology with bang and crunch, or as a crunching cosmology that evolves from a particular initial condition obtained from some quantum-cosmological tunneling event, {\it a la} Hartle--Hawking \refs\hartleh.

 At $\rho$=0 the dS$_d$ sections become null and they may be further extended as   the nearly null ${\bf H}^d$ sections of the FRW patch.  By mimicking the pure AdS case, we can achieve this matching  by
the coordinate redefinition $\rho=i\te$ and $y={\tau}+i\pi/2$: 
\eqn\hypo{
ds^2_{\rm FRW} = -d{\rm t}^{\;2} + G({\rm t})\,\left(dy^2 + \sinh^2 (y) \,d\Omega_{d-1}^2 \right) = -d{\rm t}^{\,2} + G({\rm t})\,ds^2_{{\bf H}^d}\;,
}
where the smooth matching requires  $G(\te)\approx \te^2 \approx -F (i\te)$ near $\te=0$. For the rest of the fields,  $\varphi(\rho)$ continues to a $O(1, d\,)$-invariant function $\varphi({\rm t})$ with small $\te$ behavior $\varphi(\te) \approx \varphi_0 - \shalf \varphi''_0 \,\te^2$. Hence, the result is a FRW model
with negative spatial curvature \crunchin, which  eventually  crunches barring fine-tuning.

 \bigskip
\centerline{\epsfxsize=0.6\hsize\epsfbox{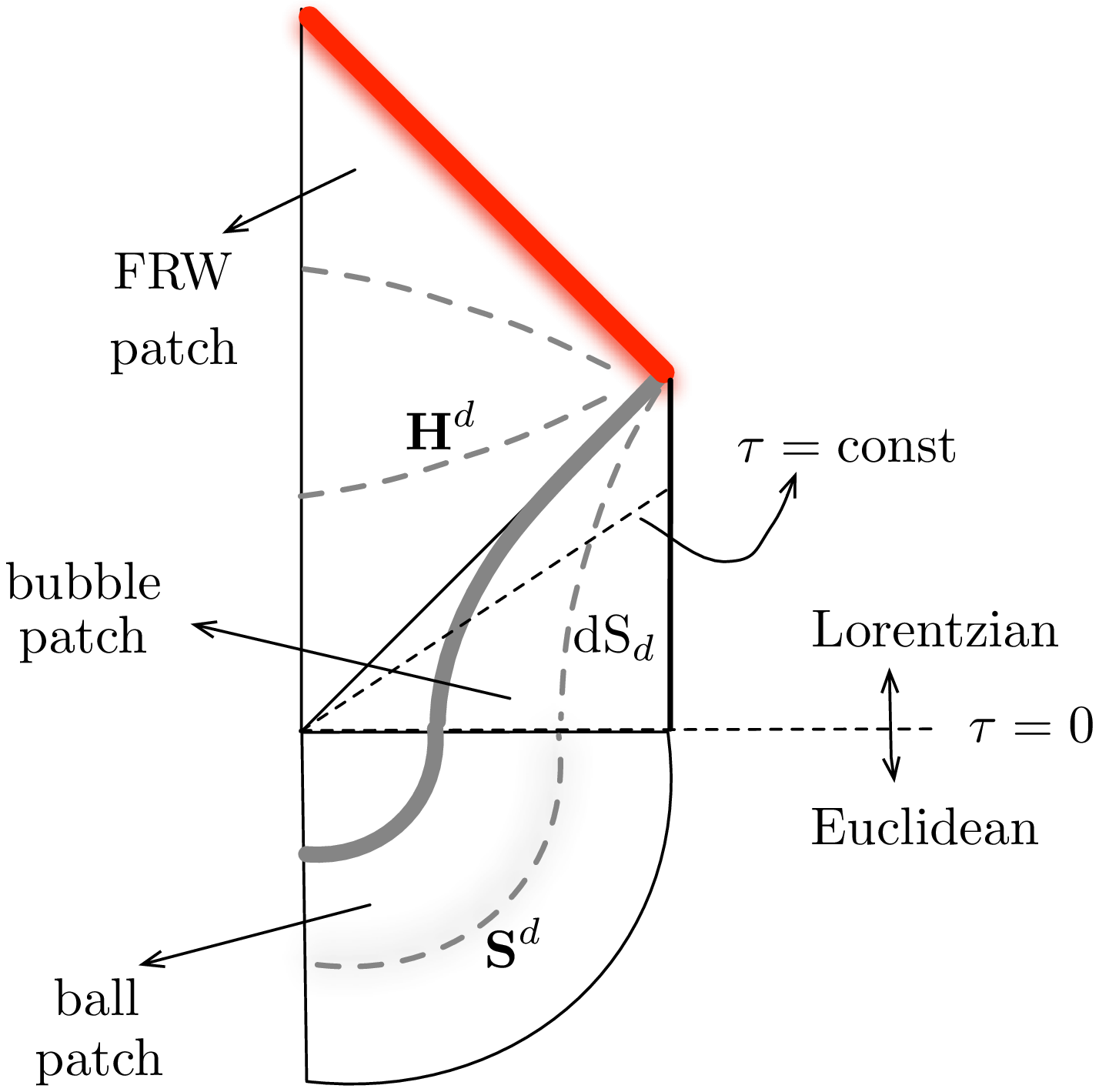}}
\noindent{\ninepoint\sl \baselineskip=8pt {\bf Figure 1:} {\ninerm
Schematic diagram showing the analytic construction of the crunching FRW space-time, with ${\bf H}^d$ spatial sections, matching to the `bubble' patch with dS$_d$ time-like sections, and finally the Euclidean `ball' background, with ${\bf S}^d$ sections. The Euclidean piece is $O(d+1)$-invariant, whereas the Lorentz-signature piece is $O(d,1)$-invariant.  The grey line signals the radius $\rho=\rho_\CO \sim \log \Lambda_\CO$,  where the metric begins to differ significantly from the asymptotic AdS form. Equivalently, it is associated to the energy scale $\Lambda_\CO$ in the dual QFT.}}
\bigskip

\subsec{The duals}

\noindent

We now restrict further the form of the ball metric to asymptote AdS,  $F(\rho) \rightarrow \sinh^2 (\rho)$ as $\rho\rightarrow\infty$. Using the Euclidean AdS/CFT rules \refs\adscft, this background describes a certain large-$N$ master field for  a perturbed dual CFT on the Euclidean $d$-sphere ${\bf S}^d$: 
 \eqn\accesfera{
\int_{{\bf S}^d} \CL_{\rm QFT} = \int_{{\bf S}^d} \CL_{\rm CFT} + \sum_\CO \int_{{\bf S}^d} {g_\CO \over (\Lambda_\CO)^{\Delta_\CO -d}}  \,\CO
\;,}
where  $\Delta_\CO$ is the conformal dimension of the perturbing operator and the dimensionless source terms $g_\CO$ are determined by the boundary values of the different fields, $\varphi_\CO$,  according to the rule
\eqn\mattb{
 g_\CO = \lim_{\rho\to\infty} \left(e^{-\rho} \, \Lambda_\CO\right)^{\Delta_\CO -d} \,\varphi_\CO (\rho)\;
, }
with $\Delta_\CO$ the conformal dimension of the operator $\CO$ and $\Lambda_\CO$ the energy scale set by the   operator perturbation. All bulk dimensions are measured in units of the asymptotic AdS radius of curvature, and all CFT length dimensions are measured in units of the  ${\bf S}^d$ radius. 

For the case considered here, with $O(d+1)$ symmetry, the $g_\CO$ are actually constant on ${\bf S}^d$ and therefore represent couplings rather than sources. Consistency at the level of the field equations requires that  only the couplings associated to relevant or marginal operators may be non-vanishing, since otherwise the back-reaction would destroy the assumed AdS asymptotic behavior of the metric. This condition can actually be extended to include also marginal operators, i.e.  we have  $g_\CO =0$ for $\Delta_\CO \geq d$ with no loss of generality, since a marginal perturbation can be conventionally folded into the definition of  $ \CL_{\rm CFT}$.  

Using the  UV/IR correspondence, 
scales $\Lambda_\CO \gg 1$ are associated to  characteristic geometric features at radii $\rho_\CO \sim \log (\Lambda_\CO)$, and thus we may speak of the radial development of the gravity solution as depicting a certain renormalization-group flow.  If the theory has no other large dimensionless parameters,   we may encounter a number of different qualitative scenarios according to the overall geometrical features of the solution:

\item
{(1)} The  flow   may become singular inside the ball at a radius of order $\rho_\CO$. In some cases the singularity  
 admits a geometric resolution, for example by a vanishing cycle in an additional compact factor of the space-time, producing  a `confining hole' at the center of the ball. The Lorentzian versions of such models would define confining gauge theories on de Sitter, examples of which can be found in \refs\Marolf\ and references therein.  Irrespectively  of   an eventual resolution of the singularity, any such background which does not reach smoothly $\rho=0$  cannot be Wick-rotated  into an FRW patch and thus does  not provide crunch models. 

\item
{(2)}
A flow with $\Lambda_\CO \gg 1$ may evolve into an approximate IR fixed point at the origin of the ball, thus depicting a flow between a UV fixed point CFT$_+$ and a IR fixed point CFT$_-$ . This corresponds to a `thin-walled bubble' of AdS$_-$ inside AdS$_+$ with radius ${\bar \rho} \sim \log(\Lambda_\CO)$, so that the bulk fields at the $\rho=0$  origin of the ball  come close to a local minimum of the supergravity potential with  
$m^2_\varphi ({\rm AdS}_-) >0$.

\item{(3)}
The flow below the threshold $\Lambda_\CO$   may continue without  approaching an IR fixed point or a gap before it reaches $\rho=0$. In this case we have roughly a bubble with `thick walls', whose interior differs significantly from AdS, but may still define $O(d,1)$-invariant crunches by the procedure outlined above, provided the background is smooth at the origin. \foot{The so-called `thin-wall' approximation refers to the idealization in which the bubble's shell is regarded as having zero thickness. In this paper, we will refer to a `thin-walled bubble' whenever  the interior contains an approximate AdS$_-$ metric, and a `thick-walled bubble' corresponds to a bubble with no recognizable AdS metric in its interior. This terminology is meant to give a qualitative description of the bubble, rather than a strict implementation of the `thin-wall approximation'.} An interesting case of a `thick-walled' bubble is the situation with $\Lambda_\CO \ll 1$, where
the deformation is so small compared to the scale of the ${\bf S}^d$ sphere that the `bubble' is but a small `hump' around $\rho=0$ (cf. the appendix of \refs\MaldacenaUN). A well-studied class of solutions of this form can be found in \refs\HertogH, using a truncated four-dimensional supergravity model with one scalar field close to a local BF-stable \refs\bt\ maximum of the potential, $-d^2 /4 < m_\varphi^2 ({\rm AdS}_+) \leq 0$. The resulting solutions are constructed in global coordinates and admit two dual  state interpretations in a CFT which is  deformed  by either marginal (cf. \refs\HertogH) or relevant operators (cf. \refs\MaldacenaUN), depending on the choice of standard or alternative quantization for the bulk scalar. The physics of both alternative interpretations is very different, as the marginal deformation was found to trigger an instability of the  CFT on ${\rm dS}_d$ or on ${\bf S}^d$, whereas the relevant case is believed to be locally stable.

\bigskip

In all previous cases, smoothness of metric and fields implies that, generically, relevant and/or marginal operators are turned on at the UV fixed point. However, in special situations we may have `flows' which are normalizable at the boundary, so that no operators, relevant or marginal, are turned on. For example, any smooth flow which starts at the boundary from a local minimum  of the supergravity potential, $m^2_{\varphi} ({\rm AdS}_+) >0$ must necessarily be a normalizable deformation of pure AdS, because any amount of non-normalizable deformation back-reacts strongly at $\rho=\infty$ and destroys the AdS asymptotic behavior.  These normalizable flows are the CdL bounces, which  break the conformal symmetry spontaneously, and thus come in continuous families (moduli spaces) associated to bulk translations of the $O(d+1)$-invariant solution. In addition, there are dilute limits with (at least) approximate multi-bubble solutions. 

In our conventions, we regard a marginal coupling as part of the unperturbed CFT. Therefore, any non-normalizable master field which induces a marginal operator in the CFT Lagrangian may be regarded as a normalizable master field of the deformed CFT. This means that  flows associated to marginal operators can  be conventionally treated as  CdL-type backgrounds.
The defining feature of the CdL instantons is their normalizable nature at the outer boundary, irrespectively of their detailed structure in the interior, i.e. we may have CdL instantons realized as `bubbles of nothing' \refs\nothing,  as non-geometric impurities, or as smooth bubble-like backgrounds. It is the last class that allows us to study crunches in their real-time development, although they only have a putative `true vacuum' inside for the case of thin walls (see \refs\banks\ for a  tour around the various pitfalls of the theory of CdL tunneling).

The Wick rotation of any of these Euclidean `ball metrics'  (i.e.  master fields) into the bubble space-time defines  a large-$N$ {\it state}  in  the same QFT  formulated on the Wick rotation  of the ${\bf S}^d$ slices, i.e. the de Sitter slices. We have the real-time QFT on dS$_d$ with action
\eqn\accds{
\int_{{\rm dS}_d} \CL_{\rm QFT} = \int_{{\rm dS}_d} \CL_{\rm CFT} - \sum_{\CO_{\rm relevant}} \int_{{\rm dS}_d} {g_\CO \over (\Lambda_\CO)^{\Delta_\CO -d}}  \,\CO
\;.}
Geometrical features at radii  $ \rho_\CO \sim \log(\Lambda_\CO) \gg 1$ correspond to
fixed energy scales for the QFT on de Sitter space-time, measured in units of the Hubble constant of dS$_d$. 
For instance, we may have a  confining theory in de Sitter space, which corresponds in the bulk to an expanding `bubble of nothing' (in this case associated to a non-normalizable deformation of asymptotic AdS). A case of more interest for the purpose of discussing crunches is the state obtained by 
Wick rotation of the `domain wall flow', which defines a state looking like the vacuum of a CFT$_+$ in the UV, and as the vacuum of a CFT$_-$ in the IR. In the bulk, we simply see a thin-walled bubble of AdS$_-$ expanding exponentially into the external AdS$_+$.  

In this last situation, there is an alternative CFT representation of the bubble interior, in terms of the IR CFT fixed point: we just write the same expression in \accds\  replacing the UV CFT$_+$ by the IR CFT$_-$, 
\eqn\accdsir{
\int_{{\rm dS}_d} \CL_{\rm QFT_-} = \int_{{\rm dS}_d} \CL_{\rm CFT_-} - \sum_{\CO_{\rm irrelevant}} \int_{{\rm dS}_d} {g_\CO \over (\Lambda_\CO)^{\Delta_\CO -d}}  \,\CO
\;,}
and  restricting the sum over perturbing operators to the infinite tower of {\it irrelevant} operators, according to the operator content of  the CFT$_-$ fixed point (cf. \refs\MaldacenaUN). This gives an approximate Wilsonian description of the bubble's interior with UV cutoff $\Lambda_\CO$, after the bubble's wall and whatever lies outside has been `integrated out'. In particular, the bubble could be sitting
inside an asymptotic $(d+1)$-dimensional de Sitter or Minkowski space-time, and the IR description would be very similar, provided the AdS$_-$ fixed point exists.  Thus, universality of the Wilsonian flow around an IR fixed point explains the fact that all nearly-AdS crunching FRW cosmologies look roughly the same, irrespectively of the initial conditions. The down side of this Wilsonian description is that details of the UV completion are hidden in
the properties of the full tower of irrelevant operators in \accdsir. 

The crunch singularities of the FRW patch start at the boundary and `propagate inwards' into the bulk. Therefore, the crunch makes its first appearance in the deep UV regime of the dual QFT,  and its properties are potentially very sensitive to the details of the UV completion. 

\newsec{Facing the CFT crunch time is complementary}

\noindent

The Lagrangian  \accds\ gives a seemingly   well-defined holographic representation of  the bubble patch, i.e. a de Sitter-invariant state of a deformed CFT in  de Sitter space-time.  The Lagrangian \accdsir\ gives also a well-defined, albeit {\it approximate}, description of the bubble quantum mechanics in the case that
an approximate AdS$_-$ interior exists. Either of these `CFT-on-dS' pictures have a causal patch in the bulk which leaves out the crunch, as it happens `after' the end of de Sitter time $\tau=\infty$. On the other hand, the $\tau =0$ surface of the bubble patch is a valid Cauchy surface for the complete  crunching manifold in the bulk. This means  that  the $\tau=0$ state of the theory \accds\ {\it does} contain all the relevant information to probe the crunch. In order to expose this information, we need a `complementarity transformation' which in this case is essentially determined by the symmetries of the problem. 

To find the complementarity map, notice that an `infalling' observer is characterized by meeting the crunch in finite time. In standard black hole states, only sufficiently infrared probes have the chance to fall into the black hole (i.e. thermalize)  and  become eligible `to meet the crunch'. This is actually a complication for the problem of finding the complementarity map in the QFT, since it entangles it with the renormalization group. In the case of the AdS cosmological crunches, however, the singularity is visible from the boundary, and thus it can be probed by arbitrarily UV states of the theory. So one strategy is to simply find a time variable in the boundary which, unlike de Sitter time, sees the crunch coming in finite time.

  Any dS$_d$ slice at constant $\rho$ corresponds to an accelerating, asymptotically null trajectory which hits the AdS$_+$ boundary in finite static time.  This suggests that we should use static AdS time in order to describe states in such a way that  they  `fall' into the crunch in finite time (i.e. infalling observers). Static time is defined as the time coordinate, $t$,  adapted to the asymptotic time-like  Killing vector $\pt/\pt t$ in AdS$_+$. This Killing vector exists with a good approximation in the `exterior' of the bubble wall trajectory. Hence, we may parametrize the near-boundary
metric of the deformed AdS in a neighborhood of $t=\tau=0$ as 
\eqn\adsein{
ds^2_{\rm Bubble} \approx -dt^2 \,(1+ r^2 ) + {dr^2 \over 1+ r^2 } + r^2\,d\Omega_{d-1}^2\;.
}
This metric defines states looking in the UV  like the vacuum of a CFT
on the Einstein space ${\rm E}_d = {\bf R} \times {\bf S}^{d-1}$. Let the metric on the Einstein space be given by
$
ds^2_{\rm E} = -dt^2 + d\Omega_{d-1}^2
$, 
which is conformally related to the dS$_d$ metric as 
\eqn\conf{
ds^2_{\rm dS} = \Omega^2 (t) \, ds^2_{\rm E}\;, \qquad \Omega(t) = \cosh(\tau) ={1\over \cos (t)}\;,
}
where $t = \int \Omega^{-1} (\tau) d\tau = 2 \tan^{-1}\left[ \tanh(\tau/2)\right]$. The so-defined conformal transformation maps the `eternity' of de Sitter time into a finite interval of Einstein-time, and thus the associated Hamiltonian  `meets'  the crunch in finite $t$ time. The extension beyond the interval $-\pi/2 < t< \pi/2$ is not guaranteed however, since the conformal transformation is singular at $t=\pm t_\star = \pm \pi/2$, the Weyl function $\Omega(t)$ having a simple pole there.  Nevertheless, the transformation is a well-defined symmetry in the domain of definition of $\Omega(t)$, and we may study the physical behavior of the theory at the edges. \foot{ Conformal maps interpreted as `complementarity transformations' appear in  \refs\insightfull, under a slightly different guise.}

 \bigskip
\centerline{\epsfxsize=0.5\hsize\epsfbox{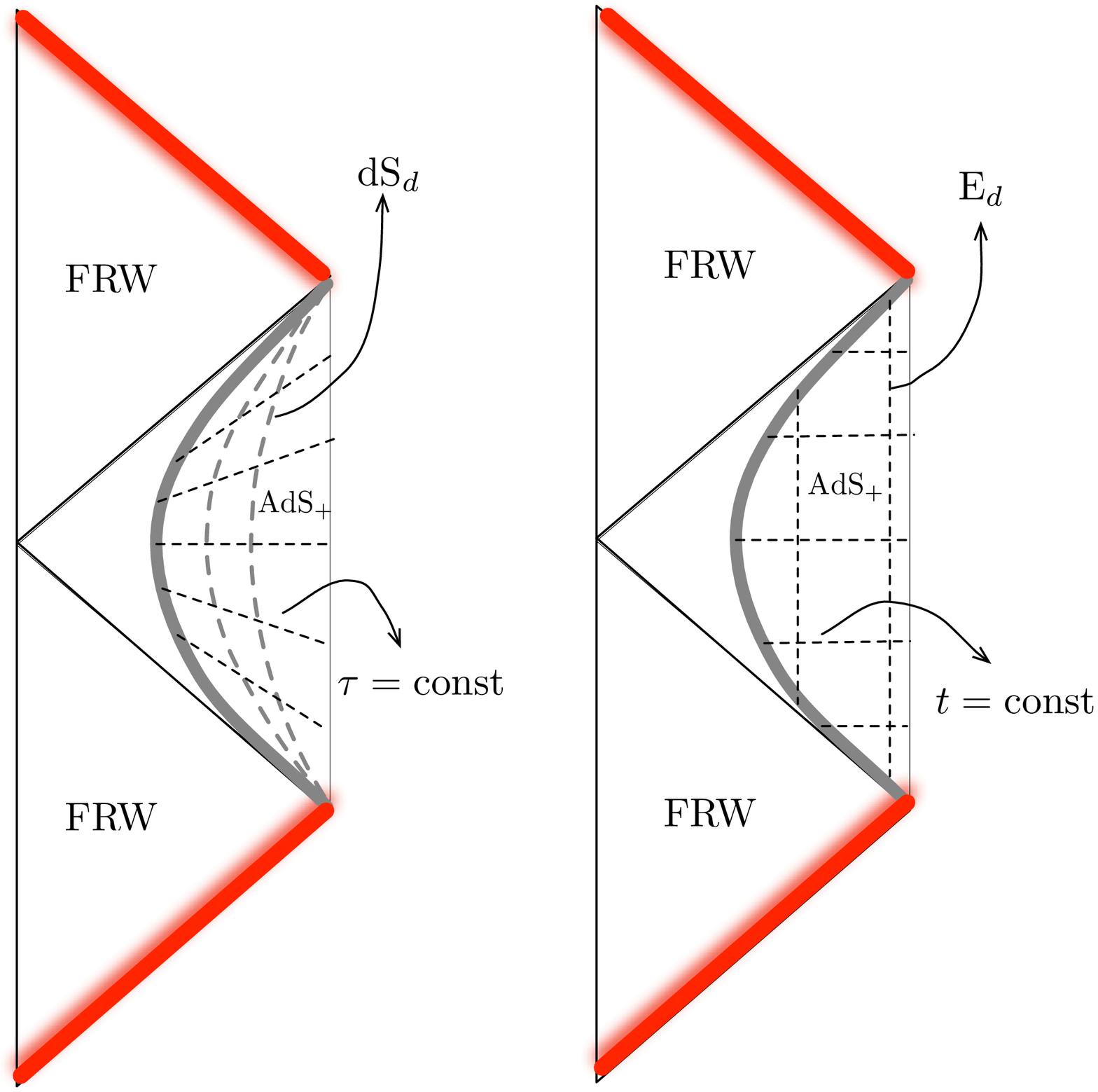}}
\noindent{\ninepoint\sl \baselineskip=8pt {\bf Figure 2:} {\ninerm
The AdS$_+$ region of the bubble patch, coordinated  along the  $O(d,1)$-invariant dS-frame (left) and the $U(1) \times O(d)$-invariant E-frame (right).  The crunch and its corresponding bang by time-reversal symmetry   are separated by an infinite amount of dS-frame time, $\tau$. In E-frame time, their time separation is finite, $\Delta t = \pi$.  }}
\bigskip

\subsec{Crashing the CFT by irresponsible driving}

\noindent

 We can now rewrite the model \accds\ in the Einstein-frame by acting with the conformal transformation \conf. Since a given operator  
transforms as  $\CO \rightarrow \Omega(t)^{-\Delta_\CO} \,\CO$  we find that the model \accds\ at {\it fixed} values of the couplings $g_\CO$ can be equivalently rewritten on the Einstein manifold as \foot{In this discussion, we neglect  the effect of conformal anomalies, a subject of great interest for future work.}
\eqn\accein{
\int_{{\rm E}_d} \CL_{\rm QFT} =\int_{{\rm E}_d} \CL_{\rm CFT} - \sum_{\CO_{\rm relevant}} \int_{{\rm E}_d} J_\CO (t) \,\CO
}
where 
\eqn\sour{
J_\CO (t) =  {g_\CO \over  \left(\Lambda_\CO (t)\right)^{\Delta_\CO -d}}\;, \qquad \Lambda_\CO (t) = \Omega(t)\,\Lambda_\CO\;.
}
If only marginal operators are turned on, this transformation is a symmetry of the deformed CFT and the Einstein-frame (E--frame) Lagrangian has no explicit time-dependence. On the other hand, each relevant operator that is turned on at the UV fixed point CFT$_+$  breaks the conformal symmetry and that is reflected on the E-frame theory as an explicit time-dependent `driving' term proportional to $J_\CO (t) \CO$.  Notice that $\Omega(t)$
diverges as $t\rightarrow t_\star$ and, since $d-\Delta_\CO >0$ for a relevant operator, the E-frame driving term, $J_\CO(t)$  `blows up' in finite time.  This conclusion holds for any $O(d,1)$-invariant and {\it relevant} perturbation of the de Sitter field theory (such time-dependent deformations were invoked in \refs{\harlow, \hsus} to avert the occurrence of multi-bubble solutions). So far this analysis leaves out the case of CdL bubbles, which include the case of an exactly marginal deformation, and will be studied in the next section.

We now argue that the value of the relevant deformation, for example  whether $g_\CO$ may be `sufficiently  positive' or `sufficiently negative', does have
a bearing on the physical interpretation in terms of a crunching cosmology. We shall give two heurisitic arguments in this section, and another one  in the next section, showing that crunching backgrounds are associated to CFT deformations contributing a sufficiently {\it negative} potential energy. 

In order to gain some intuition about such relevant driving terms, we can look at a simple model of a {\it classical}  conformal field theory with similar physical phenomena to those described here.   Consider the $O(N)$ sigma  model defined on a four-dimensional de Sitter space of unit Hubble constant, and perturbed by a  mass operator 
\eqn\landauds{
\int_{{\rm dS}_4}\CL_{\vec \phi}= -\int_{{\rm dS}_4}\left(\half \pt_\mu\, {\vec \phi} \cdot \pt^\mu\, {\vec \phi} +  {\vec\phi}^{\;2}+ g_4 \left({\vec \phi}^{\;2}\right)^2 +g_2\, \Lambda^2  {\vec \phi}^{\;2}\right)
\;,}
where we set $\Lambda \gg 1$ in order to be able to  neglect the Hubble contribution to the mass and let $g_4>0$ ensure the global  stability of the model. 
 The mass operator ${\vec \phi}^{\,2} $ is relevant but non-leading for large values of ${\vec \phi}$. 
For a positive-definite mass perturbation, $g_2 >0$, the theory acquires a mass gap of order $\Lambda$ and has a trivial IR limit. For a negative-definite mass  perturbation, $g_2 <0$, the  minimum of the potential is found at  the vacuum condensate $|{\vec \phi}\,|_{\rm vac} \propto \Lambda$ and its  low-lying excitations  are the familiar Goldstone bosons with $N-1$ degrees of freedom, the slow de Sitter expansion introducing just small thermal corrections to these vacuum states.\foot{The negative energy density at the $g_2 <0$ vacuum does not mean that we lose de Sitter, since gravity is not dynamical, i.e. Newton's constant is zero in the CFT.} Thus, this simple model with $\Lambda \gg 1$ allows us to emulate two scenarios of the classification in section 2.2. Namely it gives an example of scenario (1) for   $g_2 >0$ (a gapped phase), and a  dS-invariant domain-wall flow, or scenario (2), for $g_2 <0$. 
 
When written in the Einstein frame variables, this theory has Lagrangian
\eqn\landaue{
\int_{{\rm E}_4}\CL_{\vec \phi}= -\int_{{\rm E}_4}\left(\half  \pt_\mu \,{\vec \phi} \cdot \pt^\mu \,{\vec \phi} +\half\,{\vec\phi}^{\;2} + g_4\, \left({\vec \phi}^{\;2}\right)^2 + g_2\, \Lambda^2 (t) {\vec \phi}^{\;2}\right)
\;,}
where  now the field theory lives on a static 3-sphere but the mass operator is time dependent, with a scale $\Lambda (t) = \Lambda \,\Omega (t)$ diverging at $t=t_\star$.  

For $g_2 >0$, i.e. with no condensate in the dS description, the  theory \landaue\  exhibits an increasing gap which eventually decouples all states from the static-sphere vacuum. On the other hand, for $g_2 <0$ 
the growing tachyonic perturbation  sends
the typical  field values of the symmetry-breaking ground state to infinity,  with ever increasing  kinetic energy and  (negative) potential energy. Given the stationary `condensate'  field configuration  in the dS theory, ${\vec \phi_{\rm vac}  }$,  with $\pt_\tau {\vec \phi}_{\rm vac} =0$, the corresponding solution in the E-frame theory is                     
$\Omega(t) {\vec \phi}_{\rm vac}$, which diverges as $t\rightarrow \pm \pi/2$. 

This suggests that any {\it coherent} quantum state which is peaked around some non-zero dS-invariant configuration in the dS theory, will be mapped in the E-frame theory to a time-dependent coherent state whose support in field space is transferred to large field values. We shall refer to this transfer of `power'  in coherent states  from the IR to the UV as the `CFT fall'. 

 Notice that this happens  in the Einstein-frame formalism \landaue\ as a result of the driving term being negative-definite, even if the system  is stable at any finite value of the time variable.   On the other hand, in the de Sitter variables \landauds\ we are simply describing a certain stationary state looking like a thermal excitation of the stable symmetry-breaking vacuum.  Essentially the same physics is obtained in any model in which the formation of a condensate is controlled by the sign of a relevant operator, with a less-relevant one (not necessarily marginal) protecting the UV stability.

This example suggests that a  positive-energy driving term, which is large in Hubble units,  should be associated to 
an ever-increasing mass gap, dual to the expanding  `confinement bubble of nothing' in the bulk. Conversely, a state
with non-trivial IR content, such as one having soft degrees of freedom of an  infrared CFT$_-$, should be associated to driving operators $J_\CO (t) \,\CO$ with  a negative  contribution to the potential energy. 

In our qualitative discussion of the  $O(N)$ model we have chosen the case $\Lambda \gg 1$ to emphasize a phase structure with either a clear mass gap or a clear bosonic condensate as a ground state, in a way which is visible in the classical approximation. For $g_2 \Lambda^2 $ of the order of the Hubble scale squared, or smaller, the question of whether there is a condensate or a gap depends on the nature of the quantum corrections in the de Sitter background. For weakly coupled field theories, we do not expect a vacuum condensate to survive the thermal de Sitter bombardment for $\Lambda <1$, so that the critical value of the mass coupling, $g_2^*$, separating gapped phases from  condensates (perhaps metastable or even unstable ones), should be strictly {\it negative} and of $\CO(1)$. On the other hand, for the purposes of studying holographic duals of crunches, we are interested in semiclassical states arising as large-$N$ master fields,  and here large-$N$ effects might allow  for small-field condensates surviving the dS thermal bath, corresponding to a {\it positive} critical value of the mass deformation, $g_2^* >0$, in our toy example. 

For the particular case of the $O(N)$ model on de Sitter, the value of $g_2^*$  at large-$N$ is an interesting question that deserves further study (cf. \refs\rabroll).    
For a CFT admitting a large-$N$ gravity dual, we can get this information from the gravity solution. 
Consider for example a model like the one described in the appendix of \refs\MaldacenaUN. Here we have a 
 flow which is `stopped' by the finite-size effects of ${\bf S}^d$ before it goes non-linear, i.e. we have $\Lambda_\CO \ll 1$ or scenario (3) of section 2.2. If the gravity solution turns on a relevant boundary operator, the  field
 $\varphi(\rho)$ starts at the boundary of the ball at a local (BF-stable)  AdS$_+$ maximum of the bulk potential, which we denote conventionally $\varphi_+ =0$.   If the smooth $\varphi(\rho)$ solution  stays small throughout the `ball', $\varphi(\rho=0)$ remains close to the value of $\varphi$ at the maximum,  and it makes no difference in which direction we perturb away from $\varphi_+ =0$ (see Figure 3).

 If we now increase gradually $\Lambda_\CO$ past the inverse size of the ${\bf S}^d$ sphere, the gravity solution becomes non-linear before reaching $\rho=0$ and, in a  potential with two extrema,  such as the one depicted in Figure 3, it {\it does} make a difference in which direction we flow away from $\varphi_+ =0$. In particular, flowing to negative values of $\varphi$ we enter the basin of attraction of a AdS$_-$ minimum so that, if the initial conditions and the detailed form of the potential are just right, we may have a domain-wall solution 
 with $\Lambda_\CO \gg 1$. Conversely, flowing to positive $\varphi$ the scalar field will run away and the solution eventually develops a singularity. Since resolved `confining holes' are seen in the $(d+1)$-dimensional gravity theory as singularities, this is a typical case of a gapped phase. 
 
  \bigskip
\centerline{\epsfxsize=0.4\hsize\epsfbox{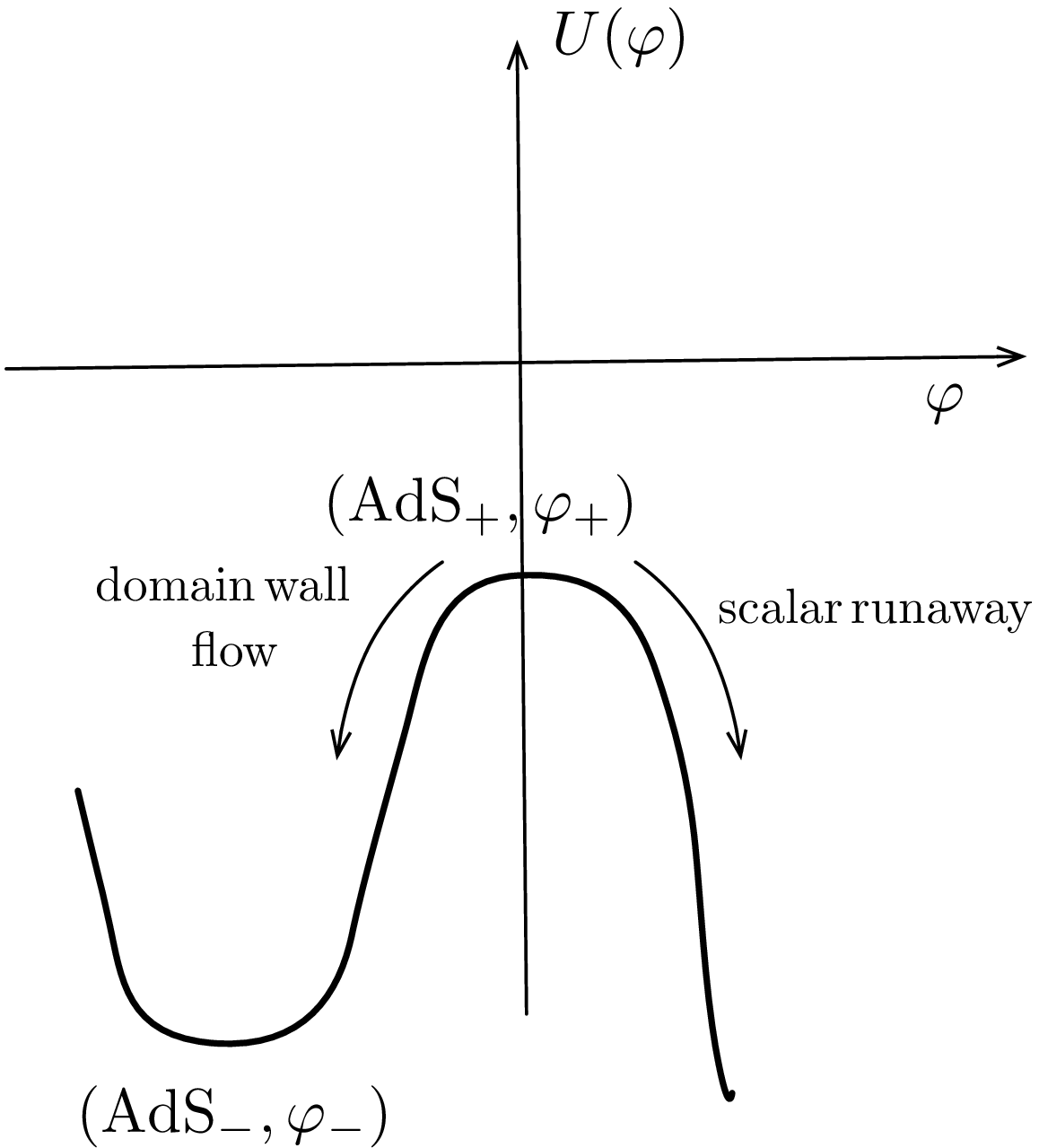}}
\noindent{\ninepoint\sl \baselineskip=8pt {\bf Figure 3:} {\ninerm
A generic bulk potential with two extrema has inequivalent flows for the two directions of $\varphi$, provided these flows enter the non-linear regime. The arrows indicate the inward evolution of the bulk scalar field $\varphi(\rho)$, starting from   an initial value $\varphi_+ =0$ at the boundary of the ball. In the figure, the  direction of negative fields may lead to the smooth domain wall flow, whereas the direction of positive fields  leads to the  scalar runaway, which generally produces  singularities in the $(d+1)$-dimensional gravity description. }}
\bigskip

 In summary, for $\Lambda_\CO \gg 1$ we recover the phenomenology of the $O(N)$-model with {\it large} $\Lambda$, suggesting that this is a generic situation, namely large, positive-definite, relevant deformations lead to gapped phases, whereas large, negative-definite, relevant deformations lead to condensates. It is  only the latter that can be used to investigate crunches. 
 
 The supergravity picture also indicates that `small', $\Lambda_\CO \ll 1$,   flows of relevant operators on ${\bf S}^d$ do generate small, stable condensates on the dS-CFT,  which can be used as holographic duals of crunches \refs\MaldacenaUN. The gravity solution ensures that this condensate exists independently of the sign of the microscopic coupling, provided the flow is sufficiently weak throughout the ball.  While the large $\Lambda_\CO$ condensates are visible in  a weakly-coupled description of the dS-CFT, the small $\Lambda_\CO$ condensates arise as a peculiar property of strongly coupled dS-CFTs admitting gravity duals.

We
shall once more return to these issues in section 4,  aided by a phenomenological  effective action, allowing us to map out the different 
scenarios of section 2.2.

\subsec{The CdL falls}

\noindent

Our interpretation of crunches as CFT falls can be made considerably sharper in the case that  we describe the bubble in the strict  thin-wall approximation. In this set up, the Euclidean background consists of  an AdS$_-$ ball of finite radius ${\bar \rho}$, surrounded by an exact AdS$_+$ metric. The physical parameters of the background are summarized by a charge $q$, controlling the jump in the AdS cosmological constant when entering the bubble,
and a tension parameter $\sigma$ for the thin shell. The condition for the Lorentzian  bubble to expand is that $0<\sigma < q$.
The same problem for a planar domain wall would yield a static solution at any value of ${\bar \rho}$ in the case
$\sigma = q$. For this reason, we refer to the condition $\sigma \geq |q|$ as a `BPS bound', despite the fact that
supersymmetry does not feature explicitly  in our analysis. With this terminology we say that only shells violating the BPS
bound correspond to ever-expanding bubbles. 

Since AdS$_+$ remains strictly unperturbed outside the bubble, this approximation describes the bubbles as normalizable states, i.e. as CdL bubbles. Therefore, the dynamics of such bubble is determined by a marginal operator in the CFT$_+$ fixed point. It was shown in \refs\us\ that this operator can be captured given the thin-wall dynamical data, i.e. the tension $\sigma$ and charge $q$ of the effective brane building the bubble's shell.

The dynamics of thin-walled vacuum bubbles, defined in terms of  junction conditions \refs{\israel,\guthblau}, can be summarized
in the effective brane action (cf. \refs\us)
\eqn\euac{
I = -\sigma\,{\rm Vol}\left[{\rm shell}\right] + d\cdot q\cdot {\rm Vol} \left[{\rm bubble}\right]\;,
}
where the first term is of Nambu--Goto type and the second term, proportional to the volume of the bubble, is of Wess--Zumino type. 

A general brane configuration can be parametrized by a collective radial field $\rho(x)$, where $x$ coordinates  the conformal boundary space-time, where the CFT is defined. Using then the results of \refs\seibwit\  and \refs\us\  we can  find the derivative expansion  of \euac\  after a convenient field-redefinition from $\rho(x)$ to  a canonically normalized  field $\phi(x)$. The leading terms for smooth and large bubbles are 
\eqn\ef{
\CL_{\rm eff} [\phi] = -\half (\pt \phi)^2 - {d-2 \over 8(d-1)} {\cal R}_d \,\phi^2 - \lambda \,\phi^{2d \over d-2} + \CO\left(\phi^{2(d-4) \over d-2}, \pt^4\right)\;.
}
where we recognize the conformal mass coupling to the background curvature and the classical marginal interaction of a conformal scalar field in $d$ dimensions, with coupling
\eqn\coupling{
\lambda = \left({d-2 \over 2}\right)^{2d \over d-2} \sigma^{2d \over 2-d} \left(1-{q \over \sigma} \right)\;.
}
Notice that, precisely for the BPS-violating case corresponding to expanding bubbles, $0<\sigma<q$, we have $\lambda <0$ and an {\it unbounded} negative potential energy fall. Hence, we confirm  that the crunch at $t=t_\star$ corresponds to the fall  down a conformally-invariant cliff which was lurking in the CFT Hamiltonian. 
 
 A crucial check  of this effective Lagrangian is  the successful matching of
 dilute instanton-gas measures,  computed in the bulk CdL description, with those corresponding to the Fubini
 instantons  \refs\fubini\ of \ef\ (cf. \refs\us, see also \refs\tassos). This matching is perfect for the instanton action, at the quantitative level, in the limit $\sigma\rightarrow q-0$.

 The Lagrangian \ef\    gives an asymptotic expansion for large $\rho (x)$, which corresponds to large $\phi(x)$ in $d\geq 2$ and to small $\phi(x)$ for $d=1$. The precise expression  in \ef\ is only valid for $d\neq 2$, the $d=2$ case being special by the occurrence of Liouville interactions. 
We should emphasize at this point that  \ef\ is not of course the full CFT Lagrangian. Rather, it is the effective Lagrangian of a particular set of configurations with the bulk geometrical interpretation of expanding shells. 
 In models with a detailed microscopic definition, such as those based on maximally supersymmetric Yang--Mills theory, the thin shells can be associated to  particular branes, such as spherical D3-branes in the AdS$_5 \times {\bf S}^5$ model  with a double-trace deformation studied in \refs\craps. In these cases, the system can  gradually jump among  $\CO(N)$ supergravity vacua by ejecting D3-branes. A similar model with different details is the same theory defined on a compact hyperboloid (cf. \refs{\insightfull, \magan}).

 As expected, we find  that the  CFT supporting a CdL bubble does have an absolutely unstable direction at large values of the collective field $\phi$. Since this unstable operator is found to be marginal, it is conformally invariant and thus  {\it also visible} in the de Sitter frame CFT. The large-bubble action of this  dS-CFT sector is obtained by writing \ef\ on dS$_d$: 
\eqn\dsac{
\int_{{\rm dS}_d} \CL_{\rm eff} [\,\phi\,] = -\int_{{\rm dS}_d} \left(\half (\pt \phi)^2 + {d(d-2) \over 8}  \,\phi^2 + \lambda \,\phi^{2d \over d-2}+ \dots \right) \;, 
}
where we have used $\CR_d = d(d-1)$ for dS$_d$. The dS-invariant solution corresponding to the thin-walled CdL bubble is $\phi_{\rm bubble} = {\bar \phi}$ with
$$
{\bar \phi} = \left({ (d-2)^2 \over 8 |\lambda|}\right)^{d-2 \over 4}\;,
$$
the constant field configuration at the maximum of the potential.  This homogeneous  field configuration coincides of course with the $O(d+1)$-invariant Euclidean solution, a local maximum of the Euclidean action, demonstrating explicitly its `bounce' character (i.e. the existence of one negative fluctuation eigenmode.)

 The  theory in Einstein-frame
variables is 
\eqn\efac{
\int_{{\rm E}_d}\CL_{\rm eff} [\,{\tilde \phi}\,] = -\int_{{\rm E}_d} \left(\half (\pt {\tilde \phi}\,)^2 + {(d-2)^2 \over 8}  \,{\tilde \phi}^{\;2} + \lambda \,{\tilde \phi}^{\;{2d \over d-2}} + \dots\right) \;,
}
with  the E-frame field  related to the dS-frame field through ${\tilde \phi}= \Omega(t)^{d-2\over 2} \,\phi$. Notice that the curvature mass is  different from that of the dS$_d$ theory, because now $\CR_d  = (d-1)(d-2)$. This slight difference of the potential means that the dS-invariant bubble solution is  not seen at the maximum of the E-frame potential, but rather corresponds to the time-dependent zero-energy trajectory 
$$
{\tilde \phi}_{\rm bubble} (t) = \Omega(t)^{d-2\over 2} \,{\bar \phi}\;,
$$
so that now ${\bar \phi}$ is the time-symmetric  $t=0$ turning point of the `CFT-fall', or the nucleation point, if we think of the real-time bubble as emerging from a pure AdS$_+$ background by a spontaneous  tunneling process \refs\us.      It is clear that any such CdL bubble can be nucleated anywhere inside AdS$_+$, so that any classical solution of \dsac\ is 
 metastable to further nucleation of Fubini bubbles,  including the $O(d,1)$-invariant configuration $\phi={\bar\phi}$  that we have  hitherto considered (cf. \refs{\us, \harlow}).

 In general, unstable operators can be expected to be marginal only in a approximate sense.  In the absence of extra degrees of freedom, the $\lambda\, \phi^{2d\over d-2}$ theory is not exactly conformally invariant, since $\lambda$ renormalizes logarithmically.\foot{This is true both in perturbation theory and in the bulk picture when $\lambda$ is associated to a double-trace operator \refs\dtrace.} Take for example the well-known case
 of $d=4$, where negative $\lambda$ is asymptotically free, leading to the corrected
 potential
 \eqn\afreedom{
 \lambda \,\phi^4 \rightarrow  \lambda(\phi)\,\phi^4 =-{\phi^4 \over \log\left(\phi/\Lambda_{\rm IR}\right)} 
 \;,
 }
 where $\Lambda_{\rm IR}$ is the RG-invariant strong-coupling scale in the infrared. Hence, under quantum corrections the fall is not exactly conformal, the instantons are not exactly defined as normalizable solutions, and only exist  in an approximate sense. The corrected background is actually marginally relevant, and we are back to the discussion in the previous section, where the driving source is given by the term \afreedom\ with
 a time-dependent IR Landau pole \foot{The time-dependent IR Landau pole $
 \Lambda_{\phi^4} (t)$ eventually becomes larger than unity at times $|t_s - t_\star | \sim (\Lambda_{\rm IR})^{-1}$,  i.e. slightly before the crunch,  the E-frame description becomes non-perturbative on the scale of the ${\bf S}^{d-1}$ sphere.}  $\Lambda_{\phi^4} (t) = \Lambda_{\rm IR} \Omega(t)$.  Incidentally, this argument gives additional evidence supporting  our prior interpretation of driven crunches as associated to negative energy falls, at least for the case of marginally relevant operators, whose effects can be approximated by those of a marginal operator.

 \newsec{Attempt at a `thin-thesis'}
 
 \noindent
 
 Given  that a marginally relevant operator can produce effects qualitatively similar to those of an exactly marginal operator, it is of interest to  pursue the dynamical description of more general bubbles in terms of effective Lagrangians of the type
 \dsac\ and \efac. In particular, let us consider bubbles in the generic situation where there are relevant operators turned on in the dual dS-CFT. If the bubbles are thin-walled, namely an approximate AdS$_-$ patch is visible in the interior, we can expect the global dynamics of the bubble to be well described by a single collective field $\phi(x)$, 
 of the type introduced in the previous subsection. Even for thick bubbles we may expect that the qualitative energetics   be accounted for by a collective field controlling the overall size and shape of a large and smooth bubble. After integrating out the rest of the degrees of freedom of
 the full dS-CFT, we can write down a phenomenological Landau--Ginzburg model  for $\phi$  whose leading operators, in a large-$\phi$ and long-distance expansion, are completely determined by conformal symmetry:
\eqn\effdsrc{
\int_{{\rm dS}_d} \CL_{\rm eff} [\,\phi\,] = -\int_{{\rm dS}_d} \left( \half (\pt \phi)^2 + {d(d-2) \over 8}  \,\phi^2   + \sum_{\Delta \leq d} \lambda_\Delta\, (\Lambda_\Delta)^{d-\Delta } \,\CO^{(\Delta)}_{\rm eff} (\phi)  + \dots \right)\;,
}
where  $\CO_{\rm eff}^{(\Delta)} (\phi) = \phi^{2\Delta \over d-2}$ is the  effective  operator of classical conformal dimension $\Delta \leq d$,  accounting for the leading  effects of  any relevant or marginal operators, $\CO_{\Delta}$,  that may be turned on at the    UV  fixed point, CFT$_+$, as in  \accds. In principle, the Lagrangian \effdsrc\ can be rigorously derived in the large-$N$ limit starting form a given supergravity background in `the ball'. The analysis of \refs\us, using the extreme thin-wall approximation, captures just  the
marginal coupling and the curvature-induced mass term. Computing systematic corrections to the extreme thin-wall approximation should yield the couplings $\lambda_\Delta$ of the conformal symmetry-breaking operators with $\Delta<d$.  

The marginal coupling $\lambda_d$ controls the extreme UV behavior, at large values of the collective field $\phi$, and thus it must be positive if the undeformed CFT$_+$ is to be absolutely stable. The example of the previous section, corresponding to CdL bubbles in the thin-wall approximation, had $\lambda_d =\lambda <0$, leaving a globally unstable direction. By allowing logarithmic effective operators  as in \afreedom\ we can also use the model to discuss marginally relevant falls (cf. \refs\craps).
    Non-generic
CFTs may have $\lambda_d=0$, like  for instance $\CN=4$ SYM projected  along the Coulomb branch. In that situation the global stability  depends on the leading less-relevant operator.  

Despite the similarity to \landauds\ and its obvious generalizations, it is important to notice that the  model \effdsrc\  has a very different status. While \landauds\ is a microscopic  (UV) 
 definition of a toy deformed CFT, we have to think of \effdsrc\ as a large-$N$  effective Lagrangian for the single  collective mode $\phi(x)$. Therefore, the classical approximation to \effdsrc\ describes all large-$N$ quantum dynamics of the underlying CFT, for the particular case of states looking like bulk bubbles.  Notice also that the effective couplings $\lambda_\Delta$ appearing in \effdsrc\ are functions of the microscopic couplings $g_\CO$ featuring in \accds\ or the toy model \landauds. However, they are related by {\it a priori} complicated dynamics and no simple relation exists between, say the sign of a given effective coupling $\lambda_\Delta$, and the analogous microscopic coupling $g_\Delta$, except perhaps when these couplings are sufficiently large, and we have a configuration which is well-approximated by a thin-walled bubble. 
 
 The essential qualitative behavior can be illustrated by a simplified  case with   a single relevant operator inducing a scale $\Lambda$,  and a single marginal operator of coupling $\lambda$:
 \eqn\effdsr{
\int_{{\rm dS}_d} \CL_{\rm eff} [\,\phi\,] = -\int_{{\rm dS}_d} \left( \half (\pt \phi)^2 + {d(d-2) \over 8}  \,\phi^2   + \lambda\,\phi^{\,{2d \over d-2}} +   \lambda_\Delta\, \Lambda^{d-\Delta } \,\phi^{\,{2\Delta \over d-2}}  + \dots \right)\;,
}
We choose $\lambda>0$ to ensure global stability of the CFT$_+$ fixed point, and we assume 
that neither $\lambda$ nor $|\lambda_\Delta |$ are   parametrically large, so that the only energy hierarchy in the system is controlled by the scale $\Lambda$.

De Sitter-invariant bubbles or equivalently, large-$N$ de Sitter-invariant states in the dS theory, are described in this approximation as extrema of the  effective potential in \effdsr. For $\lambda_\Delta \geq 0$ the only dS-invariant solution is the dS vacuum, with no scalar condensate ${\bar \phi}=0$. 
 Non-trivial condensates  require $\lambda_\Delta <0$, corresponding to a  negative-definite,  relevant perturbation in the {\it effective} theory. In this situation there are still different scenarios depending on the value of  $\Delta$  and the strength of the relevant perturbation, compared to the Hubble scale. 
 For very relevant perturbations, $\Delta < d-2$,   the only non-zero bubble solutions occur in the weak-perturbation regime,  $\Lambda \ll 1$, where we find a   local minimum determining a small dS-invariant bubble  
$
{\bar \phi}_- \sim \Lambda^{\alpha} \ll 1$, with 
$$
\alpha \equiv {(d-2)(d-\Delta) \over 2(d-2-\Delta)}
\;.$$  
This type of stable dS condensate is reminiscent of  scenario (3) in the classification of section 2.2, i.e.  it is qualitatively similar to the case
studied in the appendix of \refs\MaldacenaUN,  where the condensate is so small that it is stabilized by the
positive curvature-induced mass of dS. We remind the reader that, while this condensate requires a negative-definite effective coupling $\lambda_\Delta <0$, this could be compatible with a  small range of positive values of
the microscopic relevant coupling $g_\Delta$  (cf. the discussion in section 3.1). 

On the other hand, for 
 less relevant operators, $d>\Delta > d-2$, it is the strong-perturbation limit,   $\Lambda \gg 1$, which yields interesting solutions.  We have  a stable minimum at ${\bar \phi}_- \sim \Lambda^{d-2 \over 2}$ and a local {\it maximum} near the origin
 ${\bar \phi}_+ \sim \Lambda^\alpha \ll 1$. 
 
 Finally, for $\Delta = d-2$ both the small-$\Lambda$ minimum and the large-$\Lambda$ maximum degenerate to ${\bar\phi}=0$ in this crude approximation, whereas the large $\Lambda$ stable minimum is present  at ${\bar \phi}_- \sim \Lambda^{d-2 \over 2}$.

Any dS-invariant solution at a minimum of \effdsr\ is a stable background modeling a crunch with a well-defined status as a stationary state  in the dS-CFT. For $\Lambda\gg 1$ we have the domain-wall scenario (2), in the classification of section 2.2. For $\Lambda\ll 1$ we have seen that we can parametrize a model of type (3) in the same list.

The situation is different for the  dS-invariant states at a local maximum of the \effdsr\ potential, such as  ${\bar \phi}_+ \sim \Lambda^\alpha$ for less relevant operator perturbations. At the purely classical level this state  is interpreted as  a bulk CdL bubble. Thus, this is an example of a normalizable perturbation of a bulk background with relevant operators turned on at the boundary. The dynamical implications are entirely similar to the previous discussion of CdL bubbles associated to marginal operators, i.e. any such small-field condensate will degrade by further  uncontrolled bubble nucleations
and collisions.  

 Unlike the case of the absolutely unstable dS theories of the previous section, there is now a finite amount of potential energy available in the dS-frame potential, since the dS theory is absolutely stable for $\lambda>0$. \foot{A similar model was proposed in \refs\us\ using a negative-definite marginally relevant operator to introduce the local instability, and an exactly marginal operator of the type studied in \refs\ofer\ to stabilize the system. }
 This means that the condensate-free dS vacuum decays in this theory towards a `superheated' dS$_d$ state, at least when probed on distance scales smaller than the Hubble length (above the Hubble length,  thermalization of the scalar field fluctuations  will not be as effective). The reheating temperature is of order $\Lambda \gg 1$, much larger than the starting dS temperature. Remarkably, there are bulk descriptions of superheated dS$_d$ spaces precisely
for the static patch, in terms of bulk hyperbolic black holes (see \refs\Marolf\ and references therein for a recent discussion of these holographic duals in their de Sitter incarnation).

 \subsec{Going down your own way}
 
 \noindent

  To any classical  solution $\phi_{cl} (x)$ of \effdsr, we can associate  a classical solution  ${\tilde \phi}_{cl} (x) = \Omega(t)^{d-2 \over 2} \,\phi_{cl} (x)$ of the `driven' E-frame theory   
\eqn\effer{
\int_{{\rm E}_d} \CL_{\rm eff} [\,{\tilde \phi}\,] = -\int_{{\rm E}_d} \left(\half (\pt {\tilde \phi}\,)^2 + {(d-2)^2 \over 8}  \,{\tilde \phi}^{\,2} + \lambda\,{\tilde \phi}^{\;{2d \over d-2}}+
\lambda_\Delta \,\Lambda(t)^{d-\Delta } \,{\tilde \phi}^{\;{2\Delta \over d-2}} + \dots \right) \;, 
}
where $\Lambda (t) = \Omega (t) \Lambda$ is the by now familiar time-dependent scale that effects the `driving'. This theory has a time-dependent potential with a growing gap at the origin for $\lambda_\Delta >0$ and a  negative well becoming infinitely deep in finite time for any $\lambda_\Delta <0$.

The E-frame description of a non-trivial  dS-invariant state $\phi_{\rm bubble} = {\bar \phi}\neq 0$ is a time-dependent  bubble field
$ {\tilde \phi}_{\rm bubble} (x) = \Omega(t)^{d-2 \over 2} {\bar \phi}$ exercising what we call the `CFT fall', i.e. it goes to infinity in time $\pi/2$ from its $t=0$ turning point at   ${\bar\phi}$, having started at infinity at $t=-\pi/2$. 

While the kinetic energy of this field in the E-frame diverges as $t\rightarrow t_\star$, its E-frame potential energy density also diverges to {\it negative values}. Even the value at the turning point $t=0$ has a non-positive energy density, due to the fact that the positive mass term in \effer\ is smaller than the dS-frame mass term by an amount
$\left({d-2 \over 4}\right) {\bar \phi}^2$. 

The E-frame fall of a stable dS-invariant state looks quite different from that of a CdL-like state.  In the case of  a configuration sitting at a minimum of the dS potential, its E-frame or `infalling' representation involves a rolling field which is  `gently held' by the simultaneous fall of the  E-frame potential well (cf. Figure 4). It is interesting  to ask if this E-frame field configuration sits at the instantaneous minimum of the time-dependent E-frame potential, or rather it `rolls' to some extent, relative to the overall fall of the potential itself. The answer is that the
${\tilde \phi}_{\rm bubble}(t)$ configuration is slightly shifted from the instantaneous minimum of the driven potential, the displacement  becoming smaller as time goes by. This is again a consequence of the slight mismatch of curvature-induced masses, i.e. the factor of $d(d-2)$ versus $(d-2)^2$ between \effdsr\ and \effer, an entirely analogous effect to the more dramatic case of the marginal fall discussed in the previous section, where the potential did not acquire time dependence in the E-frame, but the field configuration did.  We can explicitly illustrate the effect in the case of a large ($\Lambda \gg 1$)  four-dimensional mass deformation, $\Delta = 2$ and $d=4$. We find
$$
{\bar \phi} = \left({\Lambda^2 -1 \over 2\lambda}\right)^{1/2}\;
$$
for the dS-frame solution. The 
  E-frame field is 
$$ 
{\tilde \phi}_{\rm bubble} (t) = \Omega (t) \, {\bar \phi}\;,
$$
while the instantaneous minimum  sits at 
$$
{\tilde \phi}_{\rm min} (t) = \left({\Omega^2 (t) \Lambda^2 - 1/2 \over 2\lambda}\right)^{1/2}\;.
$$
Hence, ${\tilde \phi}_{\rm bubble} (t)$ is slightly larger than  ${\tilde \phi}_{\rm min} (t)$, the mismatch approaching zero as $t\rightarrow t_\star$, and being maximal at $t=0$.

 \bigskip
\centerline{\epsfxsize=0.6\hsize\epsfbox{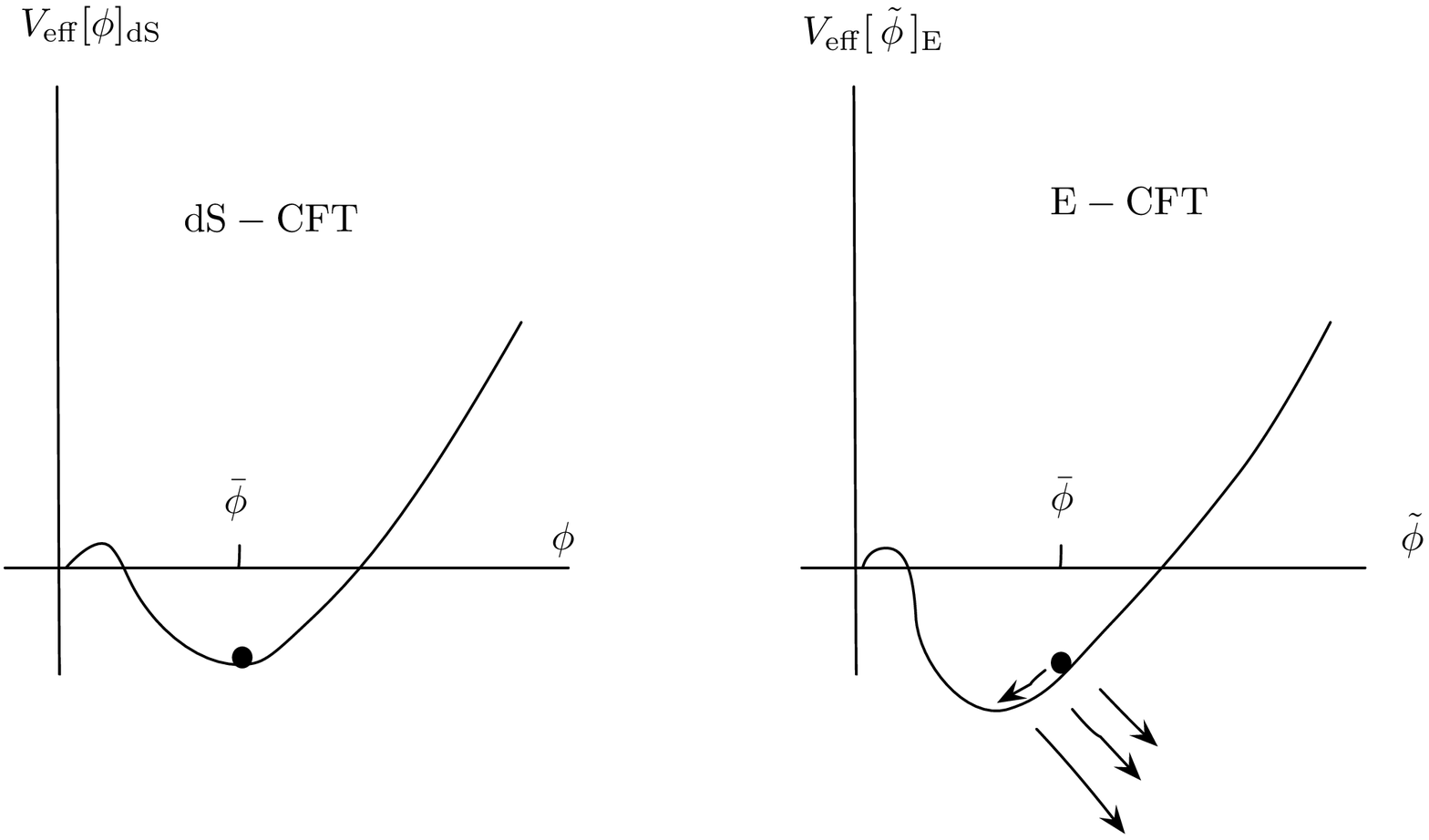}}
\noindent{\ninepoint\sl \baselineskip=8pt {\bf Figure 4:} {\ninerm
Picture a stable fall. On the left, the dS-invariant state, $\phi_{\rm bubble} ={\bar\phi}$,  is pictured as the locally stable black dot  in the dS-frame. On the right, the E-frame  state ${\tilde \phi}_{\rm bubble} (t)$ executes a combined motion: there is  a  slow roll down the E-frame potential, starting at ${\bar \phi}$ at $t=0$ and  approaching the minimum, while simultaneously the whole potential well falls down to infinity as $t\rightarrow t_\star$.}}
\bigskip

 \bigskip
\centerline{\epsfxsize=0.6\hsize\epsfbox{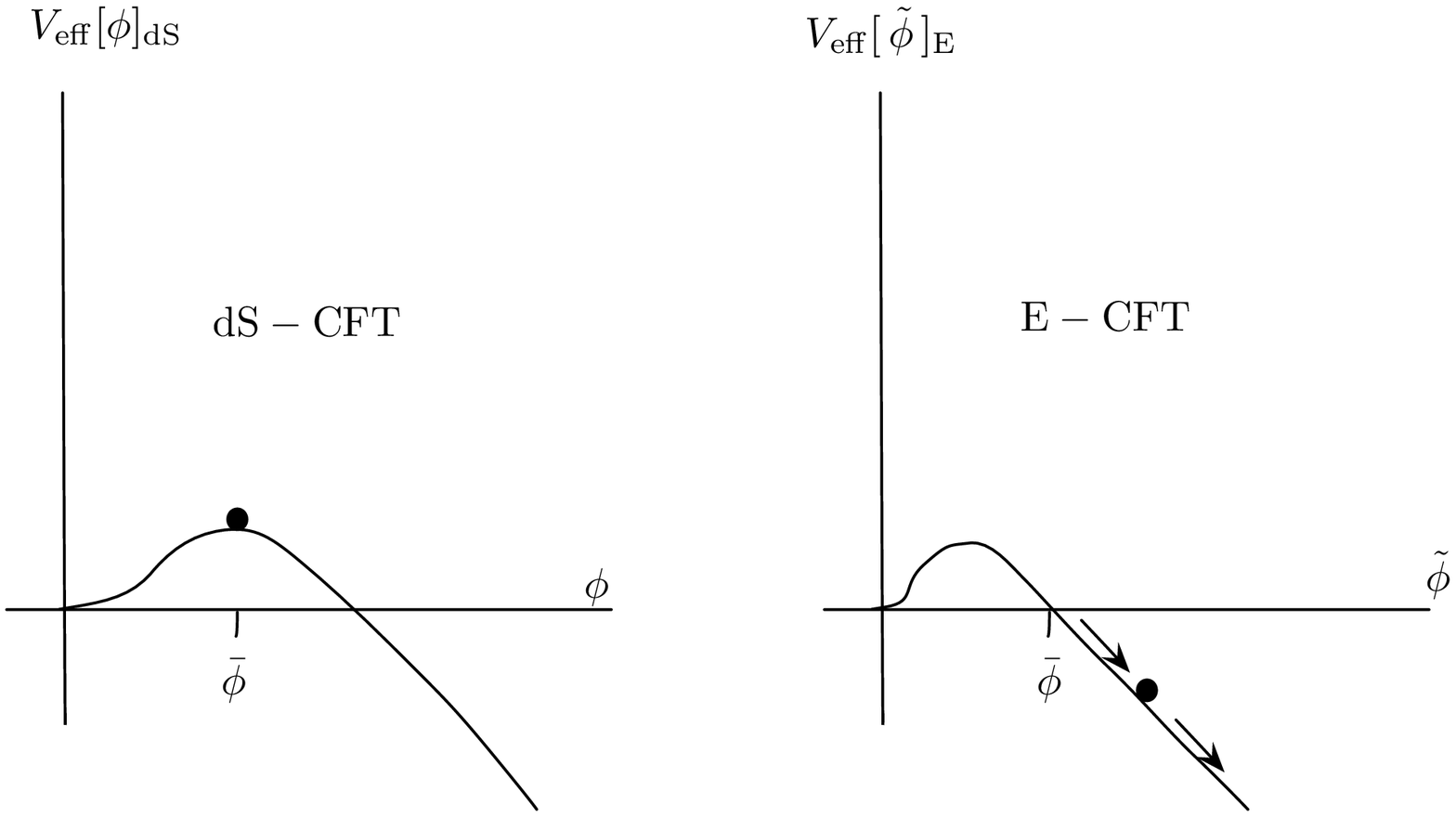}}
\noindent{\ninepoint\sl \baselineskip=8pt {\bf Figure 5:} {\ninerm
Picture of a CdL fall. The stationary   state $\phi_{\rm bubble} = {\bar \phi}$  of the dS-frame CFT sits at a local maximum (black dot on the left picture). On the E-frame CFT side we have a time-dependent falling state ${\tilde \phi}_{\rm bubble} (t)$. The potential changes slightly so that the
`sphaleron' point ${\bar \phi}$ of the dS-frame is mapped to the turning point of the E-frame configuration. If the negative slope is controlled by a marginal operator, the E-frame potential is time-independent. If the slope is controlled by a relevant operator, there is typically a receding minimum further down the potential, which may be  reached by the moving dot only at the crunch time $t=t_\star$. 
}}
\bigskip

The stated qualitative differences between the CdL-type falls and the `stable' falls refer to the behavior of  classical dS-invariant configurations at the extrema of the dS-CFT potential. We may also consider perturbations around these extrema, of the form $\phi(\tau, \Omega_\alpha) = {\bar \phi} + \delta \phi (\tau,\Omega_\alpha)$, where $\Omega_\alpha$ parametrizes ${\bf S}^{d-1}$. At the linearized level, $\delta \phi$ solves a conformally-invariant Klein--Gordon equation on dS, with long-time asymptotic behavior
$$
\delta \phi (\tau, \Omega_\alpha) \sim \exp(\gamma_\pm \tau)\;, \qquad \gamma_\pm = {1-d \over 2} \left[1\mp\sqrt{1-{4m^2 \over (d-1)^2}}\right]\;,
$$
  where $m$ is the effective mass of the linearized field $\delta \phi$. Here we see  crucial differences between the stable and unstable dS-invariant vacua. Stable vacua have   $m^2 >0$ and ${\rm Re}(\gamma_\pm) <0$, leading to decaying perturbations in the dS-frame. After transformation to the E-frame the harmonic fluctuations are still suppressed relative to the zero mode in the $t\rightarrow t_\star$ limit. In particular, the gradient energy never dominates over the kinetic energy of the field zero mode.  This means that the crunch is quite homogeneous, except for possible non-linearities generated at an intermediate time scale. 
  
  On the other hand, for the case of CdL solutions, $m^2 <0$ and ${\rm Re}(\gamma_+) >0$, so that there is an unstable solution leading to the growth of harmonic perturbations on the ${\bf S}^{d-1}$ sphere. This leads to a rapid dominance of the energy by the spatial gradient terms. Over at the E-frame the gradient modes get excited in the fall and dominate the energy density, a case of `tachyonic preheating' (see \refs\kofman\ for a review). A recent detailed analysis of such processes for a CdL fall can be found in \refs\hertogu. 
  
  These simple considerations regarding  non-homogeneous configurations show that, while crunches associated to stable CFT falls `have no hair', in the sense that the homogenous solution is linearly stable as $t\rightarrow t_\star$, crunches associated to CdL falls are quite `hairy', even at the level of single-bubble dynamics.

\subsec{Rules of engagement?}

\noindent

We finish this section with some general considerations on the overall physics picture implied by the effective Landau--Ginzburg approach.
 
One would like to identify definite field-theoretical  hallmarks for the crunch, and we have highlighted several candidates, none of which
is perfect, but each of which manifested an important aspect of the crunching phenomenon.
An indication in the E-frame was an infinite energy fall, which came in essentially two flavors. 

In one class of `CFT-fall' 
the E-frame potential is  itself  time-dependent but stable for any time before the crunch. The value of the potential at its
minimum is negative and becomes more and more negative as the time approaches the crunch time.  The field
has an expectation value  which shifts to the UV --that is, to larger and larger values. A potential in the dS-frame which
is negative at its minimum, and which has a next-to-leading operator with a large negative coupling,  would be mapped to the desired form in the E-frame. 

The other type of potential is one that is unbounded for any time in both  Einstein and de Sitter frames. It may be more palatable to accept that the crunches associated with the
stabilized potentials can exist in a theory of gravity, even if   those associated
with the second type of potentials are classically well defined for a finite time interval. \foot{Beyond the $N=\infty$ limit, which corresponds to the classical approximation to \efac, the fate of CdL fall model depends on poorly understood  multi-bubble dynamics \refs{\us, \harlow}. Even at the level of the  $1/N$ expansion, the quantum evolution of a single bubble suffers from ambiguities on   time scales arbitrarily close to $t=0$ (D. Marolf, unpublished). } 

These scenarios are particularly clear-cut when the energy scale associated to the condensate is much larger than the Hubble scale. This corresponds in the bulk with a thin-walled bubble of large size compared to the AdS radius of curvature. In this case it is natural to expect that the sign of the effective coupling, $\lambda_\CO$, is correlated with the sign of the microscopic coupling,  $g_\CO$, so that a large and negative value of $g_\CO (\Lambda_\CO)^{d-\Delta_\CO}$ is expected to be a hallmark for the crunch in the microscopic specification of the CFT. Conversely, a large and positive $g_\CO (\Lambda_\CO)^{d-\Delta_\CO}$ would be a hallmark of a gapped theory with no crunch in its bulk version.   

When the energy scale associated to the condensate is not large in Hubble units 
 the geometrical picture becomes murkier. If the slightly-negative operator is sufficiently relevant, we have found small condensates with thin walls in the Landau--Ginzburg model, mimicking  the expected behavior of supergravity solutions with bubbles which are  very small in units of the AdS curvature radius. Such supergravity solutions are, strictly speaking,  outside the realm of the thin-wall approximation and thus reducing the number of effective fields to one single collective mode  becomes questionable.
Yet it is in such circumstances that a crunch can be reliably identified in the bulk within a valid linear approximation  of the supergravity equations. This linearity implies that  there cannot be a strict correlation between the signs of $\lambda_\CO$ and $g_\CO$ in these cases. 

All examples of crunches considered in this paper seem to conform to a general rule, namely one needs a large-$N$ worth of {\it massless} degrees of freedom at the Hubble scale, coexisting with a semiclassical  condensate, which itself could be large or small in Hubble units. Expectation values  of   collective fields measuring the condensate  may then serve as  `order parameters' for the existence of the  crunch. The gapless character of the condensates seems to be essential.  Semiclassical condensates exist in gapped theories, such as gluon condensates in a confining AdS/CFT model, and yet such theories  provide unambiguous crunching models only when the confining scale is small in Hubble units, so that the glueballs still look essentially massless at the Hubble scale.

   \newsec{Falling on your sword}
 
 \noindent
 
 In this section we introduce a regularized version of the single-bubble crunching dynamics which simply makes the fall finite, further elaborating on the discussion in \refs{\HertogH,\us}. We will see that the result is a conventional black-hole final state, thus adding support to our complementarity interpretation of the map between Einstein and de Sitter frames. 
 
  From the point of view
 of the CFT in E-time frame, the crunch is associated to a finite-time fall, down an infinite cliff of negative potential energy. The geometry codifies a state in the QFT which is transferring support to UV modes, with all the characteristic energy scales diverging proportionally to  $\Omega(t)$. The infinite character of the fall is a consequence of the $O(d,1)$ symmetry of the state, since any dS invariant configuration $\phi_{\rm bubble} ={\bar \phi}$, with constant ${\bar \phi}$, is mapped to a configuration whose time dependence is just dictated by  the blowing-up Weyl factor $\Omega(t)$. Hence any regularization of the fall must break dS symmetry close to the boundary. In the dS-time frame, this means breaking the eternal character of de Sitter space-time. 
 
 When the crunch is a result of an explicit driving term, the breaking of the $O(d,1)$ symmetry must be prescribed by hand, simply declaring that the driving source $J_\CO (t)$ stops growing before the pole at $t=t_\star$. If we stop the bubble as some fixed
 radius in E-frame metric, $r_{\rm max} = \Lambda_{\rm E}$, the state breaks the $O(d,1)$ symmetry to an $U(1)\times O(d)$ symmetry, which is the global symmetry of the CFT on the Einstein manifold. In dS-frame variables, the stopping of the bubble is equivalent to terminating the `eternity' of de Sitter, with a limiting time
 $$
 \tau_{\rm max} \sim \log(\Lambda_{\rm E} / \Lambda_\CO)\;.
 $$
  The long-time  evolution in the E-frame  will take the system to a generic state with the unbroken $U(1)\times O(d)$  symmetry. If the released potential energy is large enough, those  states look like  a locally thermal state in
 the QFT, or a black hole in the bulk description. 
 
 We may estimate the energy and entropy of this final black hole in the following way. Since the driving operator $\CO$ has dimension $\Delta_\CO$, the expectation value at  long times in the E-frame is of order $\bra \CO \ket \sim N_{\rm eff} \,(\Lambda_{\rm E})^{\Delta_\CO}$, where $N_{\rm eff}$ is the central charge of the CFT, or number of effective field species in the UV limit. Therefore, the amount of potential energy released is of order
 $$
\left\bra H_{\rm driving} \right\ket \sim {g_\CO  \over (\Lambda_{\rm E})^{\Delta_\CO -d}} \,N_{\rm eff} \,(\Lambda_{\rm E})^{\Delta_\CO} \sim g_\CO \,N_{\rm eff}\,(\Lambda_{\rm E})^d 
 \;.
 $$
 If all this energy is eventually thermalized in the E-frame QFT,  it will take the form $N_{\rm eff} \,(T_{\rm eff})^d$, which
 gives an effective `reheating' temperature 
 $$
 T_{\rm eff} \sim (g_{\CO})^{1\over d} \,\Lambda_{\rm E}\;.
 $$
If $T_{\rm eff} \gg 1$, the bulk representation of this state will be a large AdS black hole of entropy
\eqn\infob{
S_{\rm BH} \sim N_{\rm eff} (T_{\rm eff})^{d-1} \sim N_{\rm eff} \,(g_\CO )^{d-1 \over d} \,(\Lambda_{\rm E})^{d-1} \;,
}
where all quantities are normalized dimensionally to the unit size of the E-frame sphere. Since $g_\CO < \CO(1)$ as part of the Wilsonian convention in defining the effective energy scales, we find that the entropy of the resulting black hole is always bounded by the maximal information capacity of the E-frame QFT, defined with a Wilsonian cutoff at scale $\Lambda_{\rm E}$,
\eqn\infobb{
S_{\rm BH} < S_{\rm max} (\Lambda_{\rm E})\;.
}

In the case of CdL bubbles, or spontaneous decay, the same arguments apply, except that now the $O(d,1)$ symmetry is broken to $U(1)\times O(d)$ by postulating a stabilization potential in  the E-frame Hamltonian, $H_{\rm CFT}$, at field values of order $\phi_{\rm max} \sim \Lambda_{\rm E}$.  For the situation studied in the previous subsection, in the thin wall approximation, the estimates about the properties of the final state apply with $g_\CO \sim | \lambda| \ll 1$ and $N_{\rm eff}$ the fraction of degrees of freedom carried by the bubble's shell, so that the information-theoretic bound \infobb\ is far from saturated. In general, the details of the approach to the typical thermalized state are much more involved here, since due attention must be paid to  multi-bubble collisions that occur within the $O(d,1)$-invariant region, $r<\Lambda_{\rm E}$.

 \bigskip
\centerline{\epsfxsize=0.30\hsize\epsfbox{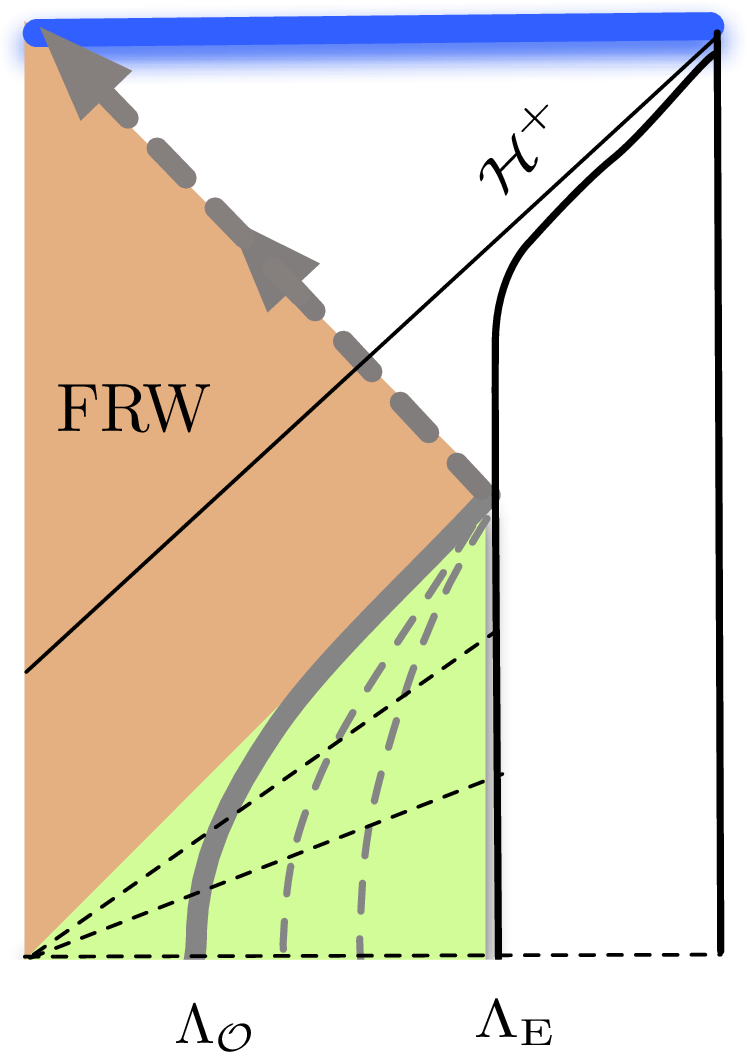}}
\noindent{\ninepoint\sl \baselineskip=8pt {\bf Figure 6:} {\ninerm
Causal diagram of the regularized crunch model. The shaded region is the part of the bulk with approximate $O(d,1)$ symmetry. The bubble patch is shown in green shade and  the FRW patch in brown. The unshaded part is the bulk region realizing the UV symmetry group $U(1)\times O(d)$. The ingoing flux of energy from the bubble collision at the UV wall bounds the FRW region and forms the final black hole.  As $\Lambda_{\rm E}  \rightarrow \infty$, this bolt of energy diverges,  as well as the black hole size, producing the FRW crunch.  }}
\bigskip

The fundamental aspect of the regularization is to regard the $O(d+1)$ group (or its Lorentzian counterpart) as an accidental symmetry emerging below a scale $\Lambda_{\rm E}$. Above this scale, the symmetry of the state is only the   $U(1)\times O(d)$ group of the Einstein space.  
Going back to the original problem posed by equation \crunchin, the resolution of the crunch by a UV breaking of the $O(d,1)$ symmetry puts the finger of blame on the large isometry of the FRW model on non-compact spatial sections. If this symmetry is broken to $U(1)\times O(d)$  outside a compact set, the crunch is tamed in the sense that it resides inside a black hole. However, it would be wrong to claim that the crunch inside the black hole is
the precise regularization of the cosmological crunch of \crunchin. A glance at the causal diagram in Figure 6 shows that the cosmological crunch is associated with the boundary of the region with $O(d,1)$-symmetry, i.e. the null surface produced by the `back flow' from the collision of the bubble with the UV wall. Inside the final black hole event horizon, this surface coincides with the apparent horizon of the black hole.

 \newsec{Conclusions}
 
 \noindent

Singularities have been encountered time and again in Physics. In most cases they marked the limits of validity of the approximation
involved. The drive to discover what lies beyond the approximation originated from the knowledge that these singularities are
not present in the phenomena described. In the study of gravity a new way emerged to understand such singularities: one considered the
possibility  that they are clocked by horizons.  This was actually realized in systems which have a holographic description. The finite
entropy that could be absorbed by a crunch inside a black hole in AdS is captured in principle by the holographic boundary observer.
In this paper we have shown explicitly how such a complementarity may work for a class of  cosmological crunches. We see that, even such singularities in
which an infinite amount of entropy flows into the crunch can be decoded by holographic observers, provided they are prepared to wait for a de Sitter eternity to complete the measurement.  We have also shown that, if in some way it turns out that  dS spaces are not eternal \refs\Polyakov,  then these singularities will involve only  finite entropy and will be regulated, as crunches in black holes are, by the system itself.

We have distilled these lessons from an analysis of certain negative-curvature FRW backgrounds with a crunching future endpoint and $O(d,1)$ symmetry. By embedding these FRW cosmologies inside expanding AdS bubbles, one can view these backgrounds as the evolution of a certain $O(d,1)$-invariant states looking like Bose condensates  of a perturbed CFT on de Sitter space-time. The {\it same} quantum state can be codified in the Hilbert space
 of {\it another} perturbed CFT defined on a static Einstein space-time. This alternative representation of the initial state uses the same basic CFT, but now perturbed by time-dependent couplings. These two descriptions of the initial state are
 related by a conformal transformation which we interpret as a `complementarity map' in the sense in which this term is used in discussions of black-hole information theory.

For those de Sitter-invariant states whose bulk development has a crunch, and the condensate has a sharply defined energy scale, we argue that its E-frame (infalling) description 
 consists on an infinite negative-energy fall, where the  quantum state
shifts coherently its support to the UV under the action of an unbounded Hamiltonian, either through an explicit time-dependent driving term or through an unbounded potential. We view this   behavior, which   we term the `CFT fall', as the CFT landmark of a crunch.  When transformed back into the dS-frame picture, the $O(d,1)$-invariant state is seen as a large-$N$ master field for a dS-CFT perturbed by {\it sufficiently}  negative-definite  relevant or marginal operators.  

As a peculiarity of the large-$N$ limit,   we can have semiclassical condensates which are  very small compared to the Hubble scale. In this case, the form of the gravity solution shows similar qualitative behavior, although it is not possible to sharply trace the `crunchy'  character to the sign of the microscopic coupling. Still, if an effective Landau--Ginzburg description 
 should exist in terms of a single collective field, it is expected to follow the general picture above.

The $O(d,1)$-invariant condensates  that harbor crunches in their bulk description fall into two broad classes, depending on whether they are  stable or unstable states in the deformed dS-CFT.  In the first case
they give self-consistent holographic models for the crunch. In the second case, corresponding to bulk backgrounds with the interpretation of CdL bubbles, the dynamics is heavily corrected by finite-$N$ effects involving multi-bubble nucleation and collisions, and even the quantum $1/N$ corrections pose a consistency challenge.

We have also investigated an explicit regularization of the CFT fall, introducing a hard UV wall with the manifest $U(1)\times O(d)$ symmetry of the E-frame theory. The $O(d,1)$ symmetry is then regarded as an accidental IR symmetry. Since the CFT fall transfers the support of the state to the UV, the reduced symmetry of the wall determines the long-time dynamics of the system, so that it approaches the generic state with $U(1)\times O(d)$ symmetry, i.e. a locally thermalized state, or a black hole in bulk parlance. 
  Within this interpretation, the FRW crunch is regularized as the null  `back flow' of debris from the collision of the CFT state with the UV wall.  Upon removal of  the UV wall,  it is the diverging energy of this back  flow what   produces the crunch. As a byproduct, this  regularization establishes a direct relation between finiteness of the black-hole entropy and the non-eternal character of the de Sitter state. At the same time, it shows that our `cosmological complementarity map' can be regarded as a limit of the more standard, and yet more mysterious,  black-hole complementarity map.

Looking at our results in the large, it is tempting to promote the scenario depicted here to the rank of a general rule, namely that the quantum description of space-like singularities is always going to involve Hamiltonian  `falls', most likely of the `stable' type, which seem well defined enough. This would apply strictly to the case in which  an infinite-dimensional Hilbert space can `fall into'  a crunch or `come from' a bang. For finite-entropy crunches/bangs one expects any continuous-time Hamiltonian description to be at best  approximate (cf. \refs\banksc).  

At this point we may ask: who is afraid of singular Hamiltonians? The `infalling' Hamiltonians described here act for a finite time and therefore {\it must} be singular. In the holographic point of view, it makes no sense to go beyond the crunch, and thus there is no need for the Hamiltonian description to be regular  at $t=t_\star$.  In any case, the examples discussed in this paper show vividly how an infinite fall propagated by an unbounded Hamiltonian can be a valid `history' of a completely regular state. 

We have one state and two non-commuting operator algebras that measure it. In one operator algebra, described as a de Sitter-CFT, the `history' is that of a stationary state, whereas in the other operator algebra, characterized as a driven Einstein-CFT, we have a singular Hamiltonian for a finite time. They are completely equivalent for the interval $-t_\star < t < t_\star$, and that is all we need because there is nothing before the bang, and nothing after the crunch, just as there is nothing left outside the eternity of de Sitter.

\bigskip{\bf Acknowledgements:}  We are indebted to T. Banks,  R. Brustein,  J. Mart\'{\i}nez--Mag\'an and S. Shenker for useful discussions. We thank the Galileo Galilei Institute for hospitality. E. Rabinovici wishes to thank Stanford University for hospitality. 
 The work of J.L.F. Barb\'on was partially supported by MEC and FEDER under a grant FPA2009-07908, the Spanish
Consolider-Ingenio 2010 Programme CPAN (CSD2007-00042) and  Comunidad Aut\'onoma de Madrid under grant HEPHACOS S2009/ESP-1473. The work of E. Rabinovici was partially supported by the Humbodlt foundation, a  DIP grant H-52, the American-Israeli Bi-National Science Foundation and the Israel Science Foundation Center of Excellence.

{\ninerm{
\listrefs
}}

\bye